\documentclass[final,5p,times,authoryear]{elsarticle}
\usepackage{amssymb}
\usepackage{amsmath}
\usepackage{array}
\usepackage{color}
\usepackage{xurl}               % smarter URL line breaks (works with hyperref)
\usepackage[hidelinks]{hyperref}
\usepackage{subcaption}
\usepackage{xcolor}

\journal{}

\begin{document}

\begin{frontmatter}

%% Title, authors and addresses

%% use the tnoteref command within \title for footnotes;
%% use the tnotetext command for theassociated footnote;
%% use the fnref command within \author or \affiliation for footnotes;
%% use the fntext command for theassociated footnote;
%% use the corref command within \author for corresponding author footnotes;
%% use the cortext command for theassociated footnote;
%% use the ead command for the email address,
%% and the form \ead[url] for the home page:
%% \title{Title\tnoteref{label1}}
%% \tnotetext[label1]{}
%% \author{Name\corref{cor1}\fnref{label2}}
%% \ead{email address}
%% \ead[url]{home page}
%% \fntext[label2]{}
%% \cortext[cor1]{}
%% \affiliation{organization={},
%%            addressline={}, 
%%            city={},
%%            postcode={}, 
%%            state={},
%%            country={}}
%% \fntext[label3]{}

%\title{The Expected Value of Mining Bitcoin} %% Article title
% \title{Risk-Return Analysis of Bitcoin Mining with and without Pooling Participation}
\title{Expected Revenue, Risk, and Grid Impact of Bitcoin Mining: A Decision-Theoretic Perspective}
%% use optional labels to link authors explicitly to addresses:
%% \author[label1,label2]{}
%% \affiliation[label1]{organization={},
%%             addressline={},
%%             city={},
%%             postcode={},
%%             state={},
%%             country={}}
%%
%% \affiliation[label2]{organization={},
%%             addressline={},
%%             city={},
%%             postcode={},
%%             state={},
%%             country={}}

\author[aff1]{Yuting Cai}
\author[aff1]{Ruthav Sadali}
\author[aff1]{Korok Ray\corref{cor1}}
\ead{korok@tamu.edu}
\author[aff1]{Chao Tian}

\cortext[cor1]{Corresponding author} %% Author name

%% Author affiliation
\affiliation[aff1]{organization={Texas A\&M University},%Department and Organization
            addressline={400 Bizzell St}, 
            city={College Station},
            state={Texas},
            postcode={77843}, 
            country={United States of America}}

%% Abstract
\begin{abstract}
%% Text of abstract
% Bitcoin mining is a lottery. Miners compute SHA-256 hash operations, and the Bitcoin protocol awards a block reward to the miner whose hash falls below a specified target. We utilize the economics of decision theory to calculate the expected value of mining bitcoin. We compare our ex ante analysis with the more common ex post numbers reported by bitcoin miners. Finally, we utilize on-chain data for the distribution of mining hardware to compute the amount of energy needed to mine a bitcoin, in expectation. This quantifies the common claim that Bitcoin is embodied energy.
Most current assessments use ex post proxies that miss uncertainty and fail to consistently capture the rapid change in bitcoin mining. We introduce a unified, ex ante statistical model that derives expected return, downside risk, and upside potential profit from the first principles of mining: Each hash is a Bernoulli trial with a Bitcoin block difficulty-based success probability. The model yields closed-form expected revenue per hash-rate unit, risk metrics in different scenarios, and upside-profit probabilities for different fleet sizes. Empirical calibration closely matches previously reported observations, yielding a unified, faithful quantification across hardware, pools, and operating conditions. This foundation enables more reliable analysis of mining impacts and behavior.
\end{abstract}

%%Graphical abstract
%%\begin{graphicalabstract}
%\includegraphics{grabs}
%%\end{graphicalabstract}

%%Research highlights
%%\begin{highlights}
%%\item Research highlight 1
%%\item Research highlight 2
%%\end{highlights}

%% Keywords
\begin{keyword}
Bitcoin mining \sep Energy consumption\sep Statistical model\sep Uncertainty quantification\sep Grid flexibility
%% keywords here, in the form: keyword \sep keyword

%% PACS codes here, in the form: \PACS code \sep code

%% MSC codes here, in the form: \MSC code \sep code
%% or \MSC[2008] code \sep code (2000 is the default)

\end{keyword}

\end{frontmatter}

%% Add \usepackage{lineno} before \begin{document} and uncomment 
%% following line to enable line numbers
%% \linenumbers

%% main text
%%

%% Use \section commands to start a section
\section{Introduction}
\label{sec1}
%% Labels are used to cross-reference an item using \ref command.
% But compared to the traditional currency, bitcoin is different because it needs to be mined out through solving cryptographic puzzles, which is also known as hash operations. In short, mining bitcoin uses computational power that costs energy for each computation operation to solve the puzzle and then receive the bitcoin reward if the puzzle is solved. Therefore, to better understand the large amount of mining behavior all over the world mentioned above and further analyze the impact of bitcoin mining on the real world, 

% Proof-of-work requires the use of energy to secure the distributed ledger, and though there have been attacks on bitcoin exchanges (like Mt. Gox), there have been no known attacks on the Bitcoin network as a whole that have disrupted the consensus enforced from proof-of-work. Since Bitcoin emerged, several alternative cryptocurrencies have proposed proof-of-stake as a way to validate the blockchain without using energy. 

% Bitcoin, the first cryptocurrency, is the largest by market capitalization and hash rate. It is secured via proof-of-work, a mechanism that requires miners to solve cryptographic puzzles to append to the blockchain. Because it works with a proof-of-work mechanism, it's hard to get disrupted and therefore become the most popular cryptocurrency worldwide.\cite{}

% To understand the impact of bitcoin mining on energy systems, we first benchmark its scale against existing electric demand. 

Bitcoin mining is unusual among industrial loads. It consumes large amounts of electricity while remaining both flexible and price-responsive. Facilities can ramp up or curtail consumption in minutes in response to prices, policy, or grid stress. This combination—high energy use and operational flexibility—creates both planning challenges and opportunities for demand side services in modern power systems.

To situate mining within energy systems, we first benchmark its scale and energy intensity. In 2017, Bitcoin was estimated to consume about 16.6~TWh/yr ($\approx$1.9~GW average demand), comparable to the annual electricity use of Slovenia \citep{krause2018quantification}. By 2022, the U.S. cryptomining load alone was approximately 5.7~GW, with about 2~GW in Texas (\cite{menati2023high}). At the hardware level, bitcoin mining is now dominated by ASIC devices, whose energy efficiency (J/TH) continues to improve, but the network-wide efficiency is determined by the active mix of new and legacy machines. 
% Operators may keep older, less efficient hardware online because capital costs are already sunk, local electricity is cheap, or new units are supply-constrained. 
Distinguishing between hardware-only network-average efficiency $\eta_e^{\text{net}}$ and facility-level efficiency $\eta_e^{\text{facility}} = \mathrm{PUE}\cdot\eta_e^{\text{net}}$ (where PUE is Power Usage Effectiveness), recent hash-rate data and plausible PUE ranges (1.01–1.20) suggest that bitcoin mining consumed on the order of $130$~TWh in 2023 and may reach $200$~TWh by 2025. At this scale, mining represents a material share of demand, and its impacts extend to operations and prices. For example, \citet{wade2025electricity} identify cryptocurrency mining as a potential driver of localized price spikes.

Although bitcoin mining poses significant challenges for power systems, it also creates opportunities. The opportunities arise in part from the scale of modern Bitcoin mining operations. For example, Riot Platforms reported 33.5 EH/s of deployed hash rate and 21.1 J/TH fleet efficiency in January 2025, corresponding to approximately 0.71 GW of mining-equipment power capacity~\citep{riot2025jan}. Comparable scales are observed among other major mining operators. CleanSpark reported 37.5 EH/s at 17.7 J/TH in December 2024, corresponding to approximately 0.66 GW~\citep{cleanspark2024dec}, while MARA reported 53.2 EH/s at 20 J/TH, corresponding to approximately 1.06 GW~\citep{mara2025dec}. To put these figures in perspective, a 1-GW load sustained continuously for one year would consume 8.76 TWh of electricity, approximately equal to the annual electricity consumption of 840,000 average U.S. residential customers~\citep{eia2025residential}. At this scale, mining operations are not marginal electricity consumers; their operating decisions can have meaningful implications for load forecasting, resource adequacy, interconnection planning, congestion, electricity prices, and local grid investment. Moreover, mining load can often be curtailed rapidly, making its response to economic and electricity-market shocks particularly relevant to grid operations.

Empirical studies show that mining load is highly price-responsive and negatively correlated with system demand at peak times~\citep{menati2023modeling,majumder2024econometric,menati2024optimization}. In practice, this combination of large scale and operational flexibility allows mines to act as controllable load resources, providing fast demand relief and ancillary services while also absorbing surplus renewable generation. These observations highlight that miners do not behave as static, inelastic loads; instead, their operations reflect deliberate economic and risk-management decisions. A central motivation of this work is, therefore, to better understand how miners respond to prices and risk, and why their behavior produces the load patterns observed in power systems.

% Empirical studies show that mining load is highly price-responsive and negatively correlated with system demand at peak times~\citep{menati2023modeling,majumder2024econometric,menati2024optimization}, indicating systematic curtailment when the grid is stressed and prices are high. In practice, large mines can participate as controllable load resources, providing fast demand relief and ancillary services that support stability and resilience, while also absorbing surplus wind and solar output and reducing renewable curtailment. These observations highlight that miners do not behave as static, inelastic loads; instead, their operations reflect deliberate economic and risk-management decisions. A central motivation of this work is, therefore, to better understand how miners respond to prices and risk, and why their behavior produces the load patterns observed in power systems.

Following \citet{menati2024optimization}, the miner–grid interaction can be viewed as a strategic game: Miners seek to maximize returns, while grid operators anticipate miner behavior and set incentive prices that recruit demand-side flexibility without burdening future grid development. To support a rigorous analysis, we must quantify the economic expectation of payoff from bitcoin mining and its exposure to the stochastic nature of block discovery. Two accounting approaches are common: ex ante (before the fact, based on assumptions or forecasts) and ex post (after the fact, based on realized outcomes); our focus is on the former to explicitly capture uncertainty and decision-dependent behavior.

% and (ii) the energy intensity (per hash) of the computation. Our ex ante probabilistic framework provides miners with a clear view of expected profit and downside risk, enabling more informed fleet sizing, pool participation, and operating strategies. At the same time, it offers system operators a behavior-aware basis for energy and capacity planning by linking miners’ potential responses to prices and risk to their resulting load profiles. 

% Following \citet{menati2024optimization}, the miner–grid interaction can be viewed as a strategic game: Miners seek to maximize returns, while grid operators anticipate miner behavior and set incentive prices that recruit demand-side flexibility without burdening future grid development. To support a rigorous analysis, we must quantify (i) the economic payoff from bitcoin mining and (ii) the energy intensity (per hash) of the computation. Two accounting approaches are common: ex ante (before the fact, based on assumptions/forecasts) and ex post (after the fact, based on realized outcomes).

Methodologically, most of the literature relies on ex post point estimates. A common proxy is the hash price—revenue per unit of computational power per day—computed from historical, point-in-time data (e.g., \citet{neumueller2025cambridge}). This approach is convenient, but deterministic. It becomes outdated quickly and does not capture the stochastic nature of bitcoin mining. In practice, block discovery is a lottery-like Poisson process. Variation in luck, difficulty retargets, and pool fees can cause realized revenue to deviate from any nominal rate. Because these ex post measures are not statistical models, they omit uncertainty and the need for miners’ risk management. From the grid’s perspective, ignoring this volatility—and miners’ dynamic responses—yields risk-unaware estimates. These can materially understate rapid swings in mining activity and load.

% Energy consumption is often estimated ex post from historical records. For example, \citet{vranken2017sustainability} inferred electricity use by dividing reported electricity expenditures by an assumed unit price. Because electricity prices vary across time and regions, this underlying method inherits high variability and can misstate actual consumption. By contrast, \citet{hayes2015cost} proposed an ex ante calculation that accounts for network difficulty, block reward, and hardware energy efficiency (e.g., W/GH). However, their setting omits facility overhead such as Power Usage Effectiveness (PUE), thereby biasing total energy downward.
% Several studies estimate energy per bitcoin as (expected hashes per BTC) $\times$ (energy per hash), where energy per hash is derived from network difficulty and hardware efficiency \citep{hayes2015cost,kristoufek2020bitcoin,vranken2017sustainability}. However, hash rate-based approaches implicitly assume homogeneous hashing effort; they underweight rapid hardware turnover, periodic difficulty adjustments, and the four-year halvings of rewards.

In short, ex post methods are incomplete because they ignore the stochastic nature of mining. We instead start from first principles—the fact that each hash is a probabilistic trial—and develop a unified, ex ante statistical framework that treats bitcoin mining as a lottery: Each hash is a Bernoulli trial with a success probability set by network difficulty, yielding a block reward upon success. This framework yields closed-form expressions for expected revenue, costs, and net profit, as well as risk and sizing rules for both direct and pool mining under different availability, fee, and risk constraints. By modeling uncertainty directly, the framework provides miners with a clear view of expected profit, downside risk, and upside potential.

% In short, ex post methods are incomplete because they ignore the stochastic nature of mining. We instead start from first principles—the fact that each hash is a probabilistic trial—and develop a unified, ex ante statistical framework that treats bitcoin mining as a lottery: Each hash is a Bernoulli trial with a success probability set by network difficulty, yielding a block reward upon success. This model provides (a) the expected revenue per unit hash rate (e.g., $\text{TH/s} \cdot \text{day}$), (b) risk metrics (variance, coefficient of variation, etc.) for solo and pool mining under varying availability, fees, and difficulty, and across risk tolerances, and (c) a consistent estimate of expected energy consumption given hardware efficiency and operating profiles. By modeling uncertainty directly, the framework guides miners’ risk-control decisions and provides grid operators with load estimates suitable for program design (e.g., demand response and ancillary-service participation).

Concretely, we conceptualize mining as follows. Each hash attempts a guess in $[0,2^{256}-1]$; the SHA-256 output is compared against a network target. If the output were below the target, the hash would “win” a block reward; otherwise, it would receive no reward. Because not every hash yields a reward, uncertainty is inherent. We show that risk can be bounded—both for direct and pool mining—via appropriate choices of participation and operating policy, using either coefficient-of-variation analysis or stochastic optimization. Under the same probabilistic framework, energy is the “ticket cost” per trial; aggregating expected trials across the network yields an internally consistent estimate of mining energy use. This unified ex ante framework serves a dual role: It helps miners design economically rational risk-control strategies and provides system operators with a quantitative tool to evaluate how bitcoin mining, treated as a controllable load, interacts with power-system operations and planning.

The rest of the paper is organized as follows. Section~\ref{sec:lottery} introduces the ex ante probabilistic model of bitcoin mining, presents its empirical calibration, and discusses the opportunity cost of mining pool participation. Section~\ref{sec:risk} applies this model to direct mining, deriving the minimum fleet size of facilities required to meet a given risk target and the probability that profit exceeds their expectations. Section~\ref{sec:risk_profit_pool} extends the same risk and upside-profit probability analysis to partial mining pool participation. Section~\ref{sec:policy} discusses the policy implications of the proposed work and Section~\ref{sec:conclusion} concludes the paper.

\section{Bitcoin mining lottery}
\label{sec:lottery}
\subsection{Mining lottery framework}
To frame bitcoin mining as a lottery, we begin with a generic binary lottery. Consider a repeated game with two outcomes: success and failure. Success occurs with probability \(p\) and yields a payoff, $R$; failure occurs with probability \(1-p\) and yields zero payoff. % Let $c$ denote the marginal cost of entering the lottery once. 
The expected value (EV) of a single play is $EV=pR$.

bitcoin mining fits this framework. A miner assembles a candidate block by selecting transactions and setting the block-header fields (including a nonce that the miner can vary). Each block is indexed by its height—its position in the chain since the genesis block. At any given height, the network specifies a difficulty target, $\tau$, which is retargeted every 2{,}016 blocks. The miner then applies the SHA-256 hash function twice to that block header; this is one hash computation. We denote the resulting 256-bit output as an integer. Operationally, each hash computation can be viewed as a single “ticket purchase” in the lottery. By this design, the output of one hash operation, $h$, is a uniform draw over $[0,2^{256}-1]$. Therefore, we can further model the output as a discrete uniform random variable with probability mass function:

% \textcolor{red}{All math notations need to be in between dollar signs, e.g., $pR>C$. }
% To explain how we can take bitcoin mining as a lottery, let's first explain how bitcoin mining works. Consider a lottery with two outcomes: success and failure. Success occurs with probability p and failure with probability 1-p. Success pays out a reward, R, and failure pays out a loss, L=0. Let c be the cost of playing the lottery. The expected value of the lottery is the weighted average of the outcomes, weighted by the probabilities, so EV=pR. This expected value must be compared to the cost of playing the lottery, c. If pR>c, then the decision maker will play the lottery; else, he will not. \textcolor{red}{This is probably incorrect. Why does a play only play the same when $pR>c$? This paragraph needs to be rewritten. What is a hash? "PDF" has not been defined?}
% For Bitcoin, we need to specify p, R, and c. To keep the analysis simple, assume the transaction fees inside a block are zero, so we focus only on the block subsidy. So, set R=6.25 BTC, the block subsidy for the 4 years before May 2024 \citep{rogers2023bitcoin}. A bitcoin miner conducts a hash operation of SHA-256. The hash operation picks a random integer h in 0,2256. \textcolor{red}{?????} Therefore, we can represent fh as the probability density function for any given hash value h. Because SHA-256 is a uniform random number generator, the PDF of a uniform distribution is simply the inverse of its support. Therefore, set the PDF of SHA-256 to be 

\begin{equation}
f(h) = \frac{1}{2^{256}}
\end{equation}

Given a network-specified target $\tau$, a hash is considered a valid proof-of-work if and only if $h<\tau$. If the miner finds such a valid hash, the miner has produced an admissible block: The network will accept it, append it to the blockchain, and credit the miner with the block reward, $R$. Because $h$ is uniformly distributed, the probability that any single hash attempt succeeds is

% To win the bitcoin lottery per block, the miner must discover a hash less than the target, which is set by the Bitcoin protocol. Therefore, the probability of winning the lottery for a given target is 

\begin{equation}
\label{eq:prob}
p(\tau) =\Pr(h < \tau)=\sum_{h=0}^{t} f(h) = \frac{\tau}{2^{256}} 
\end{equation}

We identify the payoff, $R$, with the block reward. For clarity, following industry mining reports showing that transaction fees typically account for a small share of miner revenue, e.g., 1.17\% of block rewards on average in~\citep{harper2022hashrate}, we ignore transaction fees. The expected value of one hash attempt, denominated in BTC, is therefore
% The last step follows because the SHA-256 operations are each independent random variables. Therefore, the expected value of a single hash operation is 

\begin{equation}
\label{eq:expected_value}
EV^* =p(\tau)R= \frac{\tau R}{2^{256}} \ \mathrm{BTC}/\text{Hash}
\end{equation}

% \section{Multiple Lotteries}
% \label{sec3}

% \textcolor{red}{This section needs to be rewritten.}
Bitcoin mining involves repeatedly playing this lottery. Because each hash computation is an independent trial in the lottery, the expected value of performing $H$ independent trials is simply $H$ times the expected value of a single trial. Then, the expected value of performing $H$ hash operations is

% Bitcoin mining involves repeatedly playing this lottery. We now want to calculate the expected value from multiple lotteries. The lotteries are independent, and therefore, the expected value of H-independent hash operations is 
\begin{equation}
EV_{H\_hash}=H\cdot EV^*
\end{equation}

It is convenient to define the characteristic hash operation count, $H^b$, as the number of hashes required to obtain 1 BTC in expectation. By construction,
 \begin{equation}
 \label{eq:Hstar_def}
H^b
= \frac{1}{EV^*}
= \frac{2^{256}}{\tau R}.
\end{equation}
Under this definition, $H^b \cdot EV^* = 1 \ \mathrm{BTC}$. That is, $H^b$ hashes correspond to one bitcoin in expectation, ex ante.

Let $\eta_e$ denote hardware energy efficiency (energy consumed per hash, e.g., kWh/TH).
$\eta_h$ denotes computational efficiency (hash rate in TH/s),
$p_e$ denotes the electricity price (e.g., \$/kWh), and
$T$ is the operating horizon (s).
The electricity cost over $T$ is
\begin{equation}
\label{eq:energy_cost}
  C_e \;=\; \eta_e \,\eta_h \, T \, p_e.
\end{equation}

% Let $C_h$ be the fixed one-time hardware cost allocated to the same horizon.
% Then the total cost is
% \begin{equation}
%   C \;=\; C_e + C_h.
% \end{equation}

% Let $EV_{\text{hash}}$ denote the expected bitcoin-denominated payoff per hash
% (e.g., $EV_{\text{hash}} = p(\tau)R$).
% Over horizon $T$, the expected BTC revenue is
% \begin{equation}
%   \mathbb{E}[\mathrm{Rev}_{\mathrm{BTC}}]
%   \;=\; EV_{\text{hash}} \, (\eta_h T).
% \end{equation}
If $P_{\mathrm{BTC}}$ is the BTC price in \$/BTC, the expected dollar revenue for a mining operation across $T$ seconds is:
\begin{equation}
\label{eq:fine_return}
    R_{tot}=P_{\mathrm{BTC}} \,EV^{*} \,\eta_h T   \end{equation}

In the ex post view, mining profit and cost are treated as realized and therefore deterministic. The rational rule is then evaluated only after outcomes are observed. However, this view is incomplete for planning because the number of successful hashes is unknown before mining occurs. In our ex ante framework, each hash is modeled as a probabilistic trial whose success probability is determined by network difficulty. As a result, realized revenue remains stochastic even under a fixed calibration of BTC price, electricity price, hardware efficiency, and operating horizon. Consequently, losses can still occur over finite operating periods even when expected revenue exceeds operating cost. To address this downside, we introduce explicit risk-control analysis in the following sections.

\subsection{Empirical calibration for mining lottery}
\label{sec:calibration}
% \subsection{Expected cost and return in real case}
% \label{sec:expected_cost}

% \textcolor{red}{I'm not sure about the meaning of this subsection. It seems irrelevant to the content in the previous section.  }

\begingroup
To calibrate the expected return and cost of bitcoin mining, we consider Bitcoin mainnet
block 823{,}485 (Dec 31, 2023). According to Equation (\ref{eq:expected_value}), we first 
need to identify the target, $\tau$, to determine the expected return on this block. The target is linked to the
network-reported difficulty, $D$, via
\[
D \;=\; \frac{T_1}{\tau}\quad\Longleftrightarrow\quad
\tau \;=\;  \frac{T_1}{D} ,
\]
where $T_1$ is the difficulty corresponding to the highest target, which is also known as difficulty–1. On the Bitcoin mainnet, $T_1 = 0xffff$ in hexadecimal. For block 823{,}485, the network difficulty, $D$, is approximately 72.0061 trillion \citep{coinwarz_difficulty_2025}, which is equal to $0x417C2DCA9000$ in hexadecimal

% This can be found through the difficulty, $D$, for blocks, which is a measure of how difficult it is to find a hash below a given target and is directly derived from the network. The difficulty of this block is approximately 126,271,255,279,306 \cite{}, which equals $0x72d7d150f6ca$ in hexadecimal. And the highest possible target (difficulty 1) is defined as 0x1d00ffff. 
Therefore, the $\tau$ of block 823{,}485 can be calculated as:

% First, we need to convert this decimal value to hexadecimal. This comes out to be 0x72d7d150f6ca. Rearranging the formula for difficulty:

\begin{align*}
\tau = \frac{0xffff \times 256^{0x1d-3}}{0x417C2DCA9000} \approx 3.74e^{53}.
\end{align*}
Although $\tau$ is numerically large, it remains extremely small relative to the full 256-bit space, which is why the individual success probability per hash operation remains tiny.

Substituting $\tau$ and $R=6.25 \ \text{BTC}$ \citep{rogers2023bitcoin} into Equation (\ref{eq:expected_value}) to get the expected return of a single hash attempt within this block is:

% \noindent The target comes out to: \[\tau^* = 2.13e^{54}\]

% \begin{align*}

% \tau^* = 213504928190967455231754475 \\ 471315085763075106936979456
% \end{align*}

% \noindent The new target for 2025 is less than half the size of the 2023 target. Using the previous calculation, we can calculate H* to be $1.7354853993\times10^{23}$ hashes. We can also find the expected value of a single hash operation to be:
\begin{align*}
EV^*=\frac{\tau^*(6.25)}{2^{256}} \ \text{BTC} \approx 2.02 \times 10^{-23} \ \text{BTC}/\text{Hash}.
\end{align*}

Because the hash rate in practice is quoted in terahashes per second (TH/s), it is useful to scale this expectation. One terahash (TH) is $10^{12}$ hash operations, so the expected value of one
terahash is
% \noindent The expected value of a single terahash is:
\begin{align*}
\hat{EV} = 10^{12}EV^* = 2.02 \times 10^{-11} \ \text{BTC}/\text{TH} .
\end{align*}

% In July 2025, Bitmain released the Antminer S21 XP Hyd (\citep{bitmain_s21xp_hyd_473T_12JT}). This miner has a hash efficiency $\eta_h$ of 473 TH/s\cite{}.  
Suppose that mining is performed with S19 machines, which were among the most advanced miners in 2023. This machine has a hash efficiency, $\eta_h$, of 110 TH/s \citep{bitmain_s19pro_specs}. 
% It consumes 12 J/TH.
Using this machine, and taking the bitcoin price on Dec~31,~2023 as 
$P_{\mathrm{BTC}}=\$\,42{,}265$ \citep{statmuse_bitcoin_2023}, we can find the expected return, $R_{tot}$, from mining in a year through Equation (\ref{eq:fine_return}):
\begin{align*}
R_{tot} = \underbrace{(42,265 \$/\mathrm{BTC})}_{P_{\mathrm{BTC}}}*\underbrace{(2.02\times 10^{-11}\mathrm{BTC/TH})}_{\hat{EV}}*\\
\underbrace{(110\text{TH/s} )}_{\eta_h}*\underbrace{(86,400*365\ s)}_{T}=\$2,962.95.
\end{align*}
\endgroup

Next, we compare the expected revenue with the operating electricity cost, $C_e$. The S19 machine has an energy efficiency of $29.5~\mathrm{J/TH}$ \citep{bitmain_s19pro_specs}. Converting
joules to kilowatt-hours ($1~\mathrm{kWh}=3.6\times10^{6}~\mathrm{J}$) gives
\(
\eta_e \;=\; \frac{29.5}{3.6\times10^{6}}
\;=\; 8.19\times10^{-6}\ \mathrm{kWh/TH}.
\)
With an average electricity price of \(p_e\) = \$0.0885 per kWh, the annual electricity cost for this miner over one year based on Equation (\ref{eq:energy_cost}) will be:

% \noindent This equates to \$10,173.75 in USD as of July 2025. The cost of the Antminer S21 XP Hyd is \$10,199.00. The cost of the electricity needed to mine is estimated to be \$4,395.97.
\begin{align*}
% \begin{split}
% C_e &=\frac{Energy}{Hash}\times \text{\# of Hashes/second} \\
% &\times \text{\# of seconds/year} \\ &\times \text{cost of electricity/kWh} \\
% &= 3.33 \times 10^{-6} \frac{kWh}{TH} \times 473 TH/s \\ 
% &\times \text{31,536,000 seconds} \times \\
% &\$0.0885/kWh \\
% &= \$3236.77
% \end{split}
C_{e} = \underbrace{(8.19\times 10^{-6}\text{kWh}/\mathrm{TH})}_{\eta_{e}}*\underbrace{(110~\mathrm{TH/s})}_{\eta_h}*\\
\underbrace{(86,400*365\ s)}_{T}*\underbrace{(0.0885 ~\$/\text{kWh} )}_{p_e}=\$2,514.35.
\end{align*}

For clarity, the baseline calibration treats BTC price, electricity price, network difficulty, hardware efficiency, pool payout, and transaction fees as fixed inputs. These assumptions isolate the proof-of-work lottery, which is the main source of uncertainty analyzed in this paper. The sensitivity to these inputs is direct: BTC price scales revenue, electricity price shifts net profit, network difficulty changes the success probability $(p(\tau))$, hardware efficiency affects both hash capacity and operating cost, and pool payout determines the cost of variance reduction.

\subsection{Opportunity cost of mining pools}
\label{sec:oppotunity_cost}
The analysis thus far has assumed a direct mining decision maker. Variance control under direct mining is mainly practical on a large scale; smaller facilities can instead join a mining pool to bound risk while stabilizing cash flow. A mining pool functions as an insurance intermediary: Rather than each miner facing the proof-of-work lottery, the pool aggregates participants’ hash rate, plays the lottery on their behalf, receives block rewards, and redistributes payouts according to a published scheme, net of fees. This mechanism shifts block-finding risk from the miner to the pool. However, mining pool fees and payout rules reduce expected revenue relative to direct mining. 
% We define this trade-off as the opportunity cost.
In this section, we compare the model-implied direct-mining expected output with the reported realized BTC mined by Riot Platforms and Marathon Digital in 2022. These realized gaps serve as calibration references showing the scale of the difference between ex ante expected output and realized mining outcomes in practice.

% We quantify this opportunity cost by the gap between the direct-mining expected value and the realized pool payout using the 2022 results reported by Riot Platforms and Marathon Digital, both of which participated in mining pools.

\paragraph{Riot Platforms (2022)}
In its 2022 Form 10-K, Riot reported: “as of December 31, 2022, our Bitcoin mining
business segment operated 88{,}556 miners with a hash rate capacity of 9.7~exahash per second. In 2022, we mined 5{,}554 Bitcoin” \citep{riot2022sec}. Thus, the average annual return in BTC per miner is
$5{,}554/88{,}556 \approx 0.0628~\mathrm{BTC}$.

Given 9.7~EH/s in aggregate, the average hash rate per miner is:
\[
\frac{9.7\times 10^{6}\ \mathrm{TH/s}}{88{,}556} \;\approx\; 110~\mathrm{TH/s}.
\]
Using our expected reward model, each miner is expected to earn 
\(0.0884~\mathrm{BTC}\) annually. Aggregated over 88{,}556 miners, the model yields
\(\approx 7{,}834~\mathrm{BTC/yr}\) in expectation, against the realized
\(5{,}554~\mathrm{BTC}\).
Therefore, the ex post opportunity cost for Riot Platforms in 2022 was:
% Hence the \emph{ex post shortfall} (EV minus realized) is
\[
7{,}834 \;-\; 5{,}554 \;=\; 2{,}280~\mathrm{BTC}.
\]

\paragraph{Marathon Digital (2022)}
Marathon reported \(4{,}144\) BTC mined in 2022, with \(\sim\!69{,}000\) miners and a
hash rate that increased from \(\sim\!3.5\) to \(7\)~EH/s during the year \citep{mara2022sec}.
Taking the end-of-year hash rate as a reference, the implied hash rate per miner is:
\[
\frac{7\times 10^{6}\ \mathrm{TH/s}}{69{,}000} \;\approx\; 101.45~\mathrm{TH/s},
\]
and the average BTC per miner is \(4{,}144/69{,}000 \approx 0.0601~\mathrm{BTC}\). The expected reward per miner based on this is \(0.08158~\mathrm{BTC/yr}\), implying
\(\approx 5{,}629~\mathrm{BTC/yr}\) in expectation across 69{,}000 miners, vs.\
\(4{,}144~\mathrm{BTC}\) realized. The opportunity cost is:
\[
5{,}629 \;-\; 4{,}144 \;=\; 1{,}485~\mathrm{BTC}.
\]

These magnitudes suggest that participation in the mining pool can entail significant opportunity costs. Therefore, rather than committing all machines to “risk-free” pooled mining, a practical strategy is partial pooling: Allocate a subset to the pool to secure a profit floor while leaving the remainder to reduce revenue volatility while keeping the remaining capacity exposed to direct-mining upside.
% direct mining to reduce opportunity cost and pursue higher returns.

\section{Risk control and profit upside with direct mining}
\label{sec:risk}
% \subsection{Risk control with direct mining}
As established in Section~\ref{sec:lottery}, each hash is a lottery ticket: a trial that succeeds with probability \(p(\tau)=\tau/2^{256}\), according to Equation (\ref{eq:prob}), with a payoff \(R\) or a payoff of zero. Building on this and assuming the return from each hash operation, $h$, is denoted by $\pi$, the expected return of each hash will be $\mathbb{E}[\pi]=pR$. Furthermore, the variation of the return will be $\mathrm{Var}[\pi] =\mathbb{E}[\pi^2]-(\mathbb{E}[\pi])^2$. Considering $H$ independent hashes, the expected revenue is $\mathbb{E}[H\pi]=H\mathbb{E}[\pi]=H\,p\,R$, while the variance is $\mathrm{Var}[H\pi]=H\,p(1-p)\,R^{2}$. 
% \textcolor{red}{We need to explain why, in this view, the "time period" is also an important factor.}

% To fix ideas, note that for a binary lottery \pi that rewards R with probability p and 0 with probability 1-p, expects Ex=pR and Ex2=pR2. Therefore, the variance of the lottery, X, is 

% \begin{equation}
% V[\pi]=Ex^2-(Ex)^2=pR^2(1-p)
% \end{equation}

% For n IID attempts at \pi$_i$ at this lottery, we have

% \begin{equation}
% V[\sum_{i=1}^{n} x_i]=\sum_{i=1}^{n}V[x_i]
% \end{equation}

% since the independence guarantees that the covariance is 0 between different lotteries. Using the notation from earlier, if a miner participates in the lottery H$^*$ times, then we have 

% \begin{equation}
% V[H^*\pi]=H^*V[\pi]=H^*pR^2(1-p)
% \end{equation}
Because the success probability, \(p\), of a single hash is tiny,
miners who choose to mine directly need to achieve risk control through the law of large numbers. They do this by aggregating into a larger fleet to reduce relative volatility and making cash flows more predictable. 

\subsection{Coefficient of variation risk control}
We first quantify risk with the coefficient of variation (CV), a standard dimensionless risk metric \citep{campecino2021portfolio}. CV is defined as $\frac{\sigma_\pi}{\mu_\pi}$, where $\sigma_\pi$ is the standard deviation for return of $H$ hash operations, which is equal to the square root of variation; therefore, $\sigma_\pi=R\sqrt{Hp(1-p)}$, and $\mu_\pi$ is the mean of return, which is equal to $HpR$. Assuming we want to enforce some risk control on the return, i.e., we want to control the CV below a certain threshold, $\theta$, we want $\text{CV} = \frac{\sigma_\pi}{\mu_\pi} < \theta$. After substituting $\sigma_\pi$ and $\mu_\pi$ in, we have:
\begin{equation}
\label{eq:cv_risk}
   \frac{\sigma_\pi}{\mu_\pi}=\frac{R\sqrt{Hp(1-p)}}{HpR} = \sqrt{\frac{1-p}{Hp}}<\theta 
\end{equation} 
If we want to guarantee this risk level, we need to make sure that:
\begin{equation}
\label{eq:cv_without_pool}
    H>\frac{1-p}{\theta^2p},
\end{equation}
which means we need at least $H$ hash capacity to ensure that the risk level is below $\theta$. At the facility level,
\(H \;=\; M\,\eta_h\,T,\)
where \(T_{\text{set}}\) is the settlement (accounting) horizon (calendar window for budgeting and risk),
% where $T$ is the mining horizon \textcolor{red}{This should not be called "mining horizon". It is the horizon for budgeting, accounting, or planning purposes.} 
$M$ is the number of machines, and $\eta_h$ is the average hash rate per machine. To reach the target, $H$, one facility can account for a longer time ($T \!\uparrow$), use faster hardware ($\eta_h \!\uparrow$), or deploy more machines ($M \!\uparrow$). From Equation (\ref{eq:cv_risk}), CV risk is inversely proportional to $H$; with $M$ and $\eta_h$ fixed, it is therefore inversely proportional to $T$. In other words, for a given facility, a longer mining period yields a lower risk. For risk-aware fleet planning, we fix the settlement horizon ($T=1$ year) and the hardware class (Antminer S19, $\eta_h \approx 110\,\mathrm{TH/s}$), and solve only for the required fleet size $M$ to meet the $H$ requirement in the numerical analysis below.

% \textcolor{red}{The following needs more polishing. We can perhaps mark it "example 1"?}

% \noindent\textbf{Example 1.} Assuming we are using the S19 machine with mining efficiency $\eta_h = 110 \ \text{TH/s}$ throughout the entire year, we obtain the number of machines required to reach the goal by dividing $H$ by ($\eta_h \times  (86400*365)\frac{s}{year}$). \textcolor{red}{from the formula above, it is not clear why $\eta_h$ should be in this calculation. "Dividing $H$ by...". This is confusing. Isn't the question to find out $H$? }
 
\begin{figure}[tb!]
    \centering
    \includegraphics[width=1\linewidth]{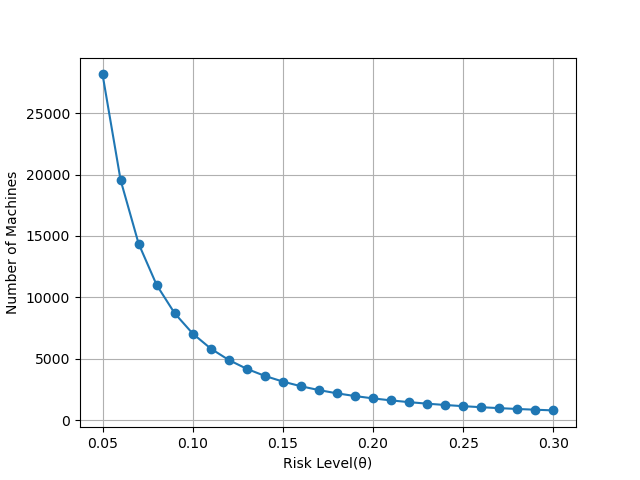}
    \caption{Minimum machines needed to meet the CV-based risk target, $\theta$.}
    \label{fig:CV_without_pool}
\end{figure}

% \textcolor{red}{we have not introduced this concept so far I believe.} \
\vspace{0.2cm}
\noindent\textbf{Example 1.} Considering the block height of 808,468 with target $\tau \approx 4.7\times 10^{53}$ \citep{coinwarz_difficulty_2025}, the corresponding success hash probability is \(p=4.067\times 10^{-24}\), according to Equation (\ref{eq:prob}). 
% and reward $R = 6.25$ during 2023.
The minimum number of S-19 type machines, $M$, required on an annual scale to achieve different levels of CV-based risk target, $\theta$, is shown in Fig.~\ref{fig:CV_without_pool}. We can observe that the number of machines increases
% \textcolor{red}{This is clearly not exponential!} 
as the risk target is set higher (i.e., the risk level is lower). For example, to maintain the risk level $\theta = 0.05$, a facility requires 28,171 machines, but only requires 3,130 machines if the risk target is set to $\theta = 0.15$. 
% \textcolor{red}{I found this discussion confusing. From the formula above, we just need $p$ and $\theta$. Why are all these parameters/variables introduced here?}

% \begin{center}
% \begin{tabular}{ |l|l| }
%  \hline
%   &  \# Machines \\ 
%  \hline
%  $\theta = 0.05 (low risk)$ & 28171 \\ 
%  \hline
%  $\theta = 0.1  (mid risk)$ & 7042 \\ 
%  \hline
%  $\theta = 0.15 (high risk)$ & 3130 \\ 
%  \hline
% \end{tabular}
% \end{center}

% \textcolor{red}{Need a new section of subsection}
\subsection{$\beta$-quantile risk control and up-side profit probability}
From a stochastic-optimization perspective, we can alternatively seek the minimal hash attempts, $H$, such that the probability of realized revenue falls below the $\alpha\in(0,1)$ profit boundary while its expectation is at risk level $\beta\in(0,1)$.

For $H$ independent hash operations, the total return is $\pi=\sum_{i=1}^{H} \pi_i = RX$, where R is the reward for a successful hash operation, and the lottery is $X \sim Binomial(H,p)$. Now assume risk level as $\beta$ and the profit variance as $\alpha$. The question of how many machines are needed to guarantee the risk level, $\beta$, and the profit variance, $\alpha$, is equivalent to finding the hash rate operation required to achieve $Pr(\pi=RX<\alpha E[\pi]) < \beta$.  Since $\mathbb{E}[\pi] = HRp$, this is equal to  Pr(X<$\alpha$ Hp) <$\beta$. The problem reduces to finding the minimal $H$ to achieve the $\beta$-th percentile of the distribution $X \sim Binomial(H,p)$. 
% \textcolor{red}{Do we always assume $\alpha\leq 1$? Why?}

% For a given binomial distribution $X \sim Binomial(H,p)$, the Chernoff bound on the tail of X can be written as:

% \begin{equation}
% Pr\{X-E[\pi]\leq - \delta\}\leq e^{-2\delta^2/H}.
% \end{equation}

% Taking $\delta= (1-\alpha)E[\pi]$ and $E[\pi] = HRp$, then this bound will be 

% \begin{equation}
% Pr\{X\leq\alpha HP\}\leq e^{-2\delta^2/H}.
% \end{equation}

% To satisfy the risk control requirement, we impose $e^{-2\delta^2/H}\leq\beta\rightarrow H \geq\frac{ln(\frac{1}{\beta})}{2(1-a)^2p^2}$. 
% Therefore, to satisfy the risk level while guaranteeing the profile bound, the facility needs to have at least $\frac{ln(\frac{1}{\beta})}{2(1-a)^2p^2}$ hash rate capability. But it’s a loose bound.
% Therefore, 

As we expect a large $H$, we can use a normal approximation for the binomial distribution:

\begin{equation}
\label{eq:normal_direct}
\begin{aligned}
X\approx Y \sim \mathcal{N}(\mu_Y, \sigma_Y^2)\\ \mu_Y = Hp,\quad \sigma_Y^2=Hp&(1-p).
\end{aligned}
\end{equation}

\noindent Now, $Pr(X<aHp)\approx Pr(Y<\alpha \mu_Y)$. After standardization:
\begin{equation}
\label{eq:normal_prob_direct}
\begin{aligned}
\Pr(Y<\alpha \mu_Y)
&= \Pr\!\left(\frac{Y-\mu_Y}{\sigma_Y}<\frac{\alpha \mu_Y - \mu_Y}{\sigma_Y}\right)
= \Pr\!\left(Z< (\alpha-1)\frac{\mu_Y}{\sigma_Y}\right) \\
&= \Pr\!\left(Z < (\alpha-1)\sqrt{\frac{Hp}{1-p}}\right),
\end{aligned}
\end{equation}
where $Z\sim N(0,1)$. This can be further written as 
\begin{align}
\Pr\left(Z < (\alpha-1)\sqrt{\frac{Hp}{1-p}}\right)=\Phi\left(-(1-\alpha)\sqrt{\frac{Hp}{1-p}}\right),
\end{align} 
where $\Phi$ is the standard normal CDF. It follows that
\begin{align}
&\Pr\left(Z < (\alpha-1)\sqrt{\frac{Hp}{1-p}}\right) < \beta\notag\\
&\leftrightarrow {}\Phi\left(-(1-\alpha)\sqrt{{\frac{Hp}{1-p}}}\right)< \beta \notag\\
&\leftrightarrow {}(1-\alpha)\sqrt{{\frac{Hp}{1-p}}}>-\Phi^{-1} (\beta).
\end{align}

\begin{figure}[tb!]
    \centering
    \includegraphics[width=\linewidth]{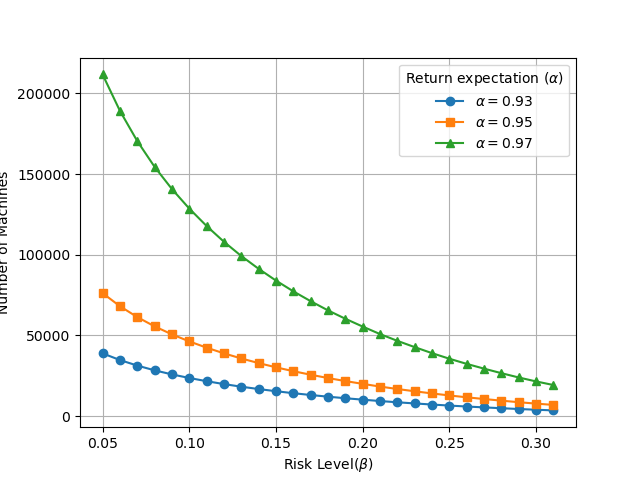}
    \caption{Minimum machines to meet risk target $\beta$ at fixed $\alpha$.}
    \label{fig:normal_without_pool_alpha}
\end{figure}

\begin{figure}[tb!]
    \centering
    \includegraphics[width=\linewidth]{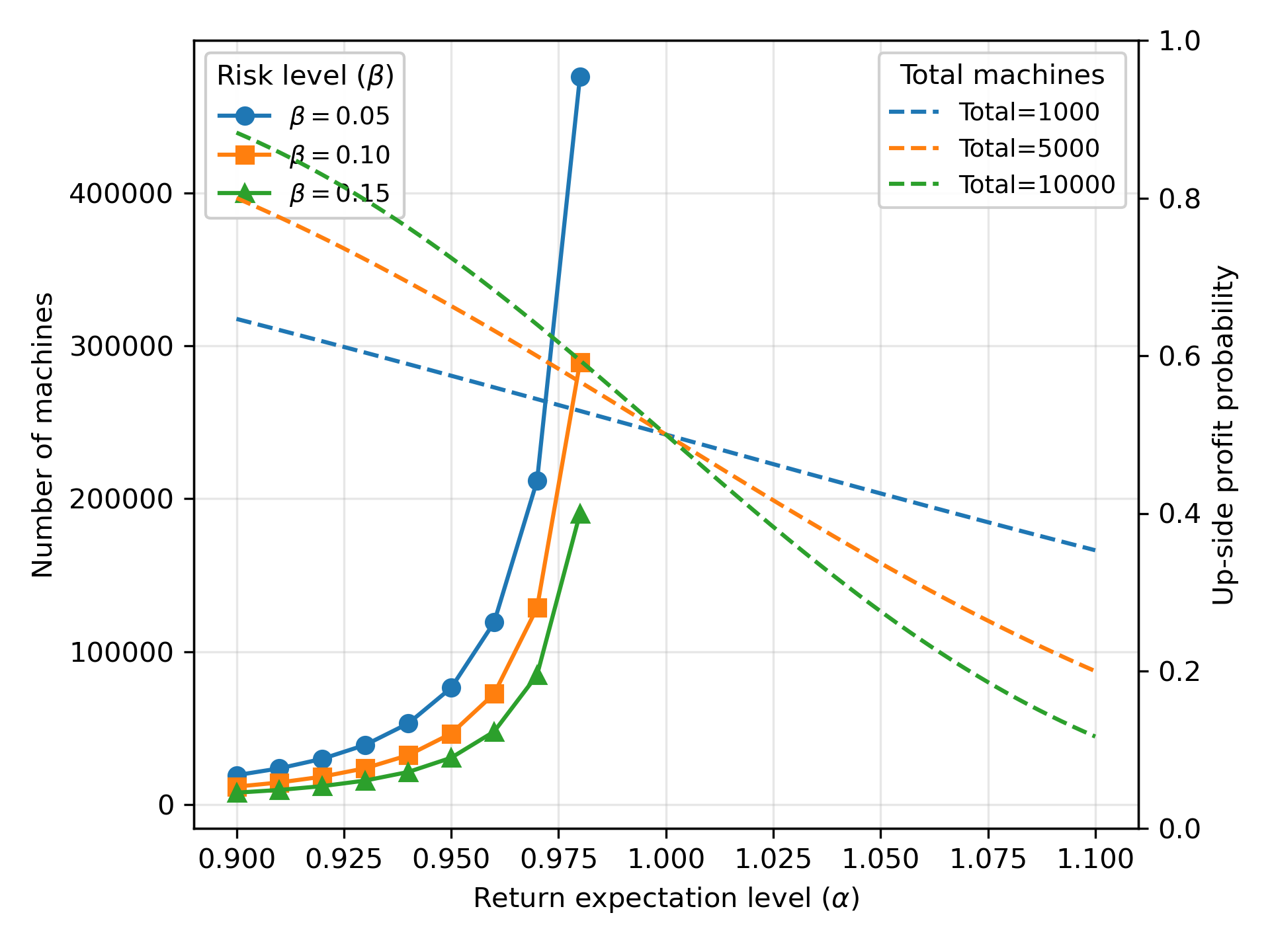}
    \caption{Minimum number of machines required to meet a relative return target $\alpha$ at risk level $\beta$ (solid lines, left axis), and probability that a fleet with total machine $M$ achieves at least $\alpha$ times its expected profit (dashed lines, right axis).}   \label{fig:normal_without_pool_beta}
\end{figure}

\noindent Given $\Phi^{-1}(\beta)=z_{\beta}$ which is the $\beta$-quantile of the normal distribution which is a constant given $\beta$, 
\begin{equation}
\label{eq:stochastic_without_pool}
\begin{aligned}
(1-\alpha)\sqrt{\frac{Hp}{1-p}} 
  > -\Phi^{-1}(\beta)&\leftrightarrow\,
  (1-\alpha)\sqrt{\frac{Hp}{1-p}} > -z_\beta \\[4pt]
&\leftrightarrow\,
  H > \frac{z_{\beta}^{2}(1-p)}{(1-\alpha)^2 p}.
\end{aligned}
\end{equation}
% \textcolor{red}{The sign changed here again $(1-\alpha)\sqrt{{\frac{Hp}{1-p}}}>\Phi^{-1}(\beta) $ for no reason! $z_{\beta}$ never appears elsewhere, and it should not be introduced.}
With \(\eta_h\) and \(T\) fixed as above, substituting \(H = M\,\eta_h\,T\) yields the minimum fleet size, \(M\), needed to meet the revenue floor, \(\alpha\), at risk tolerance, \(\beta\): Fig.~\ref{fig:normal_without_pool_alpha} reports the required \(M\) for a fixed \(\alpha\) as \(\beta\) varies; Fig.~\ref{fig:normal_without_pool_beta} reports the required \(M\) for a fixed \(\beta\) as \(\alpha\) varies. In both views, the conclusion matches the coefficient-of-variation case: Tighter risk control (smaller \(\beta\)) or a higher revenue floor (\(\alpha\)) necessitates a larger fleet, \(M\).

% Therefore, converting $H$ to $M$ by $H = M η_h T$, we obtain an approximation of the minimal type of machines, $M$, a facility should have to achieve a fixed revenue bound, $\alpha$, with different risk level, $\beta$, as shown in Fig.~\ref{fig:normal_without_pool_alpha}. \textcolor{red}{The following sentence does not read.} $M$ required to achieve a fixed risk level, $\beta$,
% with different return expectation level, $\alpha$, is shown in Fig.~\ref{fig:normal_without_pool_beta}. These results agree with the conclusion drawn from the CV case: The more restricted the risk control, the more machines are required.
% \begin{center}
% \begin{tabular}{ |l|l| }
%  \hline
%   & \# Machines \\ 
%  $\alpha = 0.95, \beta = 0.05 (low risk)$ & 76232 \\ 
%  \hline
%  $\alpha = 0.95, \beta = 0.1 (mid risk)$ & 46300 \\ 
%  \hline
%  $\alpha = 0.95, \beta = 0.15 (high risk)$ & 30470 \\ 
%  \hline
% \end{tabular}
% \end{center}

% \begin{figure}[t]
%   \centering
%     \begin{subfigure}{0.95\linewidth}
%     \includegraphics[width=\linewidth]{plot/normal_without_pool(fix_alpha).png}
%     \caption{Minimum machines to meet risk target $\beta$ at fixed $\alpha$.}
%     \label{fig:normal_without_pool_alpha}
%   \end{subfigure}

%   \vspace{4pt} % tiny gap between the two
%       \begin{subfigure}{0.95\linewidth}
%     \includegraphics[width=\linewidth]{plot/normal_without_pool(fix_beta).png}
%     \caption{Minimum machines to meet return target $\alpha$ at fixed $\beta$.}   \label{fig:normal_without_pool_beta}

%   \end{subfigure}
% \end{figure}

The risk control problem can also be expressed without approximation as an optimization problem:
\begin{equation}
\begin{aligned}
&\min_{H}\ H \\
&\text{s.t. }\ \sum_{i=1}^{\alpha Hp} \binom{H}{i} p^i(1-p)^{H-i} < \beta.
\end{aligned}
\end{equation}
% \begin{align*}
% min_{H}H
% \end{align*}
% \begin{align*}
% s.t. \sum_{i=1}^{aHp} C(H, i)p^i(1-p)^{H-i} < \beta
% \end{align*}
This optimization problem can be solved using exact search. %The optimal $H_{min}$ can be found to satisfy both risk and revenue expectation targets.

\vspace{0.2cm}
\noindent\textbf{Example 2.} If $\beta = 0.05$ and $\alpha = 0.95$, we find that the optimal $H_{min}$ is approximately $2.6e^{26}$. This is equal to the annual mining capacity of $M_{min}=74,971$ S19 machines at 110TH/s efficiency \citep{bitmain_s19pro_specs}. Table \ref{tab:normal_without_pool_opt} shows more results. This matches the conclusion of the normal approximation above.

\begin{table}[t] % or [!t], [h], etc.
  \centering
  \caption{Minimum machines to meet $(\alpha,\beta)$ targets (direct mining).}
  \label{tab:normal_without_pool_opt}
  \begin{tabular}{|l|r|}
    \hline
    & $M_{min}$ \\
    \hline
    $\alpha=0.95,\ \beta=0.05$ (low risk)  & 74{,}058 \\
    \hline
    $\alpha=0.95,\ \beta=0.10$ (mid risk)  & 44{,}585 \\
    \hline
    $\alpha=0.95,\ \beta=0.15$ (high risk) & 28{,}870 \\
    \hline
  \end{tabular}
\end{table}

The same normal approximation in Equation (\ref{eq:normal_direct}) also yields the probability that a given fleet with hash capacity \(H\) realizes at least \(\alpha\) times its expected reward \(Hp\), namely
\[
\Pr(X \ge \alpha H p),
\]
which is equivalent to
\[
\Pr(X \ge \alpha H p) = 1 - \Pr(X < \alpha H p).
\]
Combining this with the approximation in Equation (\ref{eq:normal_prob_direct})
gives
\begin{equation}
\label{net_profit_direct}
    \Pr(X \ge \alpha H p)
    \approx
    1 - \Phi\!\left((\alpha - 1)\sqrt{\frac{H p}{1 - p}}\right).
\end{equation}
Once we know the probability of achieving different levels of expected return, we know the probability of net profit because the electricity cost is deterministic, regardless of whether individual hashes succeed, and therefore can be treated as a fixed amount, $C_e$, for a given fleet size and time period.

\vspace{0.2cm}
\noindent\textbf{Example 3.}
Using the calibration results in Section~\ref{sec:calibration}, suppose the $1000$ S19 machines mining continuously for one year. According to Section~\ref{sec:calibration}, the expected annual revenue per machine is \$2{,}962.95, so the total expected revenue is
\[
1000 \times \$2{,}962.95 = \$2,962,950.
\]
Setting $\alpha = 1.1$, the probability that the realized revenue exceeds,
\[
\$2,962,950 \times 1.1 = \$3,259,245,
\]
is approximately $36.88\%$, obtained via Equation (\ref{net_profit_direct}). Furthermore, subtracting the deterministic electricity cost, $C_e$, calibrated in Section~\ref{sec:calibration}, where
\[
C_e = 1000 \times \$2{,}514.35 = \$2,514,350,
\]
the probability that the net profit exceeds
\[
\$3,259,245 - \$2,514,350 = \$448{,}600
\]
is also $36.8\%$. Additional probability results for other values of $\alpha$ and number of machines are shown in Fig.~\ref{fig:normal_without_pool_beta}.

\section{Risk control and profit upside with mining pool}
% \subsection{Risk control with mining pools}
\label{sec:risk_profit_pool}

% \textcolor{red}{I suggest breaking 3.1 and 3.2 into two sections.}
As noted in Section~\ref{sec:oppotunity_cost}, mining facilities often prefer partial pooling, hedging a portion of their hash-rate capacity to secure cash flow while leaving the remainder unpooled to pursue higher expected returns. Operationally, the key decision is the fraction of capacity to allocate (hedge) in the pool.

We will not explicitly model pool mechanics; rather, we use pooling to capture miners’ risk aversion and variance transfer. Consider a small facility with a nameplate hash rate capability \(H'\). The sizing question we ask is this: How many machines (and thus what effective hash rate capability, \(N'\), out of a total \(H'\)) are needed, under a chosen pool payout scheme and fee, to achieve a target profit floor while satisfying a specified risk constraint? We analyze this allocation problem under the same two risk measures as in the previous section: coefficient-of-variation (CV) risk and $\beta$-quantile risk. 
% \textcolor{red}{This is distracting. It should be moved to Section II.}

\subsection{Coefficient of variation risk control}

Suppose the facility allocates \(N^{\prime}\) of its total \(H^{\prime}\) hash attempts to a mining pool, leaving \(H^{\prime}-N^{\prime}\) attempts for direct mining. The return is
\begin{equation}
\pi^{\prime} \;=\; N^{\prime} R^{\prime} \;+\; R\,X^{\prime}, 
\qquad 
X^{\prime} \sim \mathrm{Binomial}\!\left(H^{\prime}-N^{\prime},\,p\right),
\end{equation}
where \(p\) is the success probability per attempt for direct mining, \(R\) is the block reward (without a pool), and \(R^{\prime}\) is the pool payout per contributed hash attempt. Intuitively, the pool shifts variance away from the miner: It guarantees that each contributed hash earns a smaller but deterministic payout $R^{\prime}$.

% Referring to the report estimation, based on the Riot report, one S19 machine working for a whole year will earn 0.62 BTC after joining the pool (details shown in Section \ref{sec:opportunity_cost}). $R'$ for each hash will be $R'= \frac{0.62BTC}{110TH/s*365 day/year*86400s/day*1e^{12}} = 1.78e^{-22} \ \text{BTC}$. 

We start with simple risk control based on the coefficient of variation (CV). $\text{CV} = \frac{\sigma}{\mu_{\pi'}}$, where $\sigma_{\pi'}$ is the standard deviation for the return $\pi'$ of $H'$ 
% \textcolor{red}{Pay attention to the difference between math and text. Here it should be $H$. This issue happens many times. } 
total hash operations and $N'$ hash operations in pool, which is equal to the square root of variance. Under this setting, $\sigma_{\pi'}=R\sqrt{(H'-N')p(1-p)}$ because only the hash rate capacity not protected by the mining pool has randomness. $\mu_{\pi'}$ is the expected return for machines not joining the pool, which is equal to $R'N'+R(H'-N')p$. Assume that we want to enforce a risk control on the return by keeping the CV below a certain threshold $\theta$, i.e., $\text{CV} = \frac{\sigma_{\pi'}}{\mu_{\pi'}} < \theta$. Substituting $\sigma_{\pi'}$ and $\mu_{\pi'}$ in, we have:

\begin{equation}
    \mathrm{CV} \;=\; \frac{\sigma_{\pi'}}{\mu_{\pi'}}
    \;=\;
    \frac{R\sqrt{(H^{\prime}-N^{\prime})\,p(1-p)}}
         {R^{\prime}N^{\prime} + R(H^{\prime}-N^{\prime})p}
    \;<\; \theta,
    \label{eq:cv}
\end{equation}
with \(0\le N^{\prime}\le H^{\prime}\) and \(R,R^{\prime},p,\theta>0\),
% Since the denominator in Equation (\ref{eq:cv}) is \(\mu_{\pi'} = (R^{\prime}-Rp)N^{\prime}+RpH^{\prime}\), 
squaring both sides gives
\begin{equation}
    \theta^{2}\bigl((R^{\prime}-Rp)N^{\prime}+RpH^{\prime}\bigr)^{2}
    \;-\;
    R^{2}p(1-p)\,(H^{\prime}-N^{\prime})
    \;>\; 0.
\end{equation}
Expanding and collecting terms in \(N^{\prime}\) yields the quadratic inequality
\begin{equation}
    a\,{N^{\prime}}^{2} \;+\; b\,N^{\prime} \;+\; c \;>\; 0,
\end{equation}
where
\begin{align*}
    a \;&=\; \theta^{2}\,(R^{\prime}-Rp)^{2},\\
    b \;&=\; 2\theta^{2}\,RpH^{\prime}\,(R^{\prime}-Rp) \;+\; R^{2}p(1-p),\\
    c \;&=\; \theta^{2}\,R^{2}p^{2}{H^{\prime}}^{2} \;-\; R^{2}p(1-p)\,H^{\prime}.
\end{align*}
Assume $a>0$ (i.e., $R'\neq Rp$). We focus on facilities that lack sufficient hash capacity to satisfy the direct mining risk criterion in Equation (\ref{eq:cv_without_pool}), so we take $H' < \frac{1-p}{\theta^{2}p}$. This condition is equivalent to $c<0$.%, implying that the smaller root is negative. 
 Consequently, the minimal $N'$ that satisfies the constraint is the larger root, clipped at zero:
\begin{equation}
    N^{\prime}_{\min}
    \;=\;
    max\left(0,\frac{-\,b \;+\; \sqrt{\,b^{2}-4ac\,}}{2a}\right).
\end{equation}

\begin{figure}[tb!]
    \centering
    \includegraphics[width=1\linewidth]{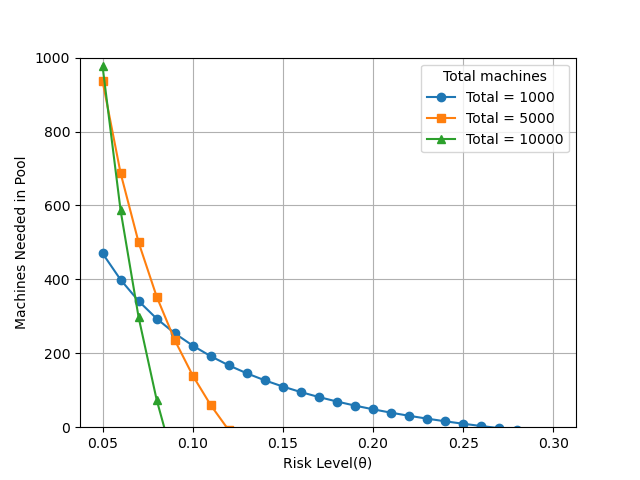}
    \caption{Minimum machines needed to join the mining pool to meet the CV-based risk target, $\theta$, for different sizes of mining facility.}
    \label{fig:CV_with_pool}
\end{figure}

As shown in Fig.~\ref{fig:CV_with_pool}, larger facilities are more resilient to return uncertainty, and therefore require a smaller pooled share to achieve a given risk target. Even smaller facilities need not commit their entire capacity to a pool. For facility-level planning, let $M'$ be the machine count corresponding to $N'$. With horizon $T$ and per-machine hash rate $\eta_h$,
\(
N' \;=\; M'\,\eta_h\,T.
\)
In this section, we fix $T=1$ year and $\eta_h \approx 110\,\mathrm{TH/s}$ (S19 class). Thus, contributing only 
\(
M' \;=\; \Big\lceil \tfrac{N'_{\min}}{\eta_h\,T} \Big\rceil
\)
machines—the amount needed to reach the minimum required exposure $N'_{\min}$—satisfies the risk constraint while limiting pool management fees and preserving higher expected returns on the remainder of the fleet.

\subsection{$\beta$-quantile risk control and up-side profit probability}
From a tail-bound perspective, fix a risk level \(\beta \in (0,1)\) (tolerated tail probability) and a shortfall factor \(\alpha \in (0,1)\).
% Let
% \[
% \pi' \;=\; N'R' \;+\; R\,X', 
% \qquad 
% X' \sim \mathrm{Binomial}\!\bigl(H'-N',\,p\bigr),
% \qquad
% \mu \;=\; \mathbb{E}[\pi'] \;=\; N'R' + R(H'-N')p .
% \]
The design question we ask is this: How much capacity should be put into the mining pool (equivalently, how many machines) to guarantee the lower-tail risk constraint? This question can be expressed as:
\begin{equation}
    \Pr\!\left(\pi' \;<\; \alpha\,\mu_{\pi'} \right) \;\le\; \beta. \
\end{equation}
Since \(\pi' = N'R' + R X'\), this is equivalent to a tail bound on \(X'\):
\begin{equation}
\begin{aligned}
\Pr \ \Bigl(
  X' &< \frac{\alpha\mu_{\pi'} - N'R'}{R}
\Bigr)
\\
&= \Pr\left( X' < \alpha(H'-N')p
      + \frac{(\alpha - 1)N'R'}{R} \right)
  \le \beta.
\end{aligned}
\end{equation}

% From a tail-bound perspective, let \(\beta\) denote the tolerated tail probability (risk level) and \(\alpha\) denote the profit variance. The question is how many machines need to be used to join the mining pool to guarantee the risk level $\beta$ and the profit variance $\alpha$? This is equivalent to find the hash rate operation required to achieve $PR(\pi'= N'R'+RX'<\alpha E[\pi]) <\beta$. Since $E[\pi] = N'R'+R(H'-N')p$, this is equal to  $PR(X'<\alpha(H'-N')p+\frac{(\alpha -1)N'R'}{R}) <\beta$. 

We can apply a normal approximation to the binomial distribution based on the hashes outside the pool, $H'-N'$, which leads to
\begin{equation}
\label{eq:normal_pool}
    \begin{aligned}
X' &\approx Y' \sim \mathcal{N}\!\bigl(\mu_{Y'},\sigma_{Y'}^2\bigr), \\
\mu_{Y'} &\coloneqq  \,(H^{\prime}-N^{\prime})\,p,\\
\sigma_{Y'}^2 &\coloneqq\,(H^{\prime}-N^{\prime})\,p(1-p).
\end{aligned}
\end{equation}
Then
\begin{equation}
\label{eq:normal_step1}
\begin{aligned}
\Pr\Bigl(
  X' &< \alpha(H'-N')p + \tfrac{(\alpha-1)N'R'}{R}
\Bigr)
\\
&\approx
\Pr\Bigl(
  Y' < \alpha(H'-N')p + \tfrac{(\alpha-1)N'R'}{R}
\Bigr).
\end{aligned}
\end{equation}
After standardization, we obtain
\begin{equation}
\label{eq:normal_step2}
\begin{aligned}
\Pr\Bigl(
  Y' &< \alpha(H'-N')p + \tfrac{(\alpha-1)N'R'}{R}
\Bigr)
\\
&=
\Pr\Biggl(
  \frac{Y'-\mu_{Y'}}{\sigma_{Y'}} <
  \frac{\alpha(H'-N')p + \tfrac{(\alpha-1)N'R'}{R}-\mu_{Y'}}{\sigma_{Y'}}
\Biggr)
\\
&=
\Pr\Biggl(
  Z <
  \frac{(\alpha-1)\bigl[(H'-N')p + \tfrac{N'R'}{R}\bigr]}
       {\sqrt{(H'-N')p(1-p)}}
\Biggr),
\end{aligned}
\end{equation}
where $Z \sim \mathcal{N}(0,1)$. Hence
\begin{equation}
\label{eq:normal_prob_pool}
\begin{aligned}
\Pr \Biggl(
  Z &<
  \frac{(\alpha-1)\bigl[(H'-N')p + \tfrac{N'R'}{R}\bigr]}
       {\sqrt{(H'-N')p(1-p)}}
\Biggr)
\\
&=
\Phi\ \Biggl(
  \frac{(\alpha-1)\bigl[(H'-N')p + \tfrac{N'R'}{R}\bigr]}
       {\sqrt{(H'-N')p(1-p)}}
\Biggr),
\end{aligned}
\end{equation}
where $\Phi$ denotes the standard normal CDF. Therefore, the approximate
risk constraint
\[
\Pr\left(
  X' < \alpha(H'-N')p + \frac{(\alpha-1)N'R'}{R}
\right) \le \beta
\]
is equivalent to
\begin{equation}
\label{eq:normal_ineq_pool}
\frac{(1-\alpha)\bigl[(H'-N')p + \tfrac{N'R'}{R}\bigr]}
     {\sqrt{(H'-N')p(1-p)}}
\;\ge\;
-\Phi^{-1}(\beta).
\end{equation}

% $Pr(Z < (1-\alpha)\frac{N'R'+R(H'-N')p}{R\sqrt{(H'-N')p(1-p)}}) < \beta \leftrightarrow{}$

% $\Phi(-(1-\alpha)\frac{N'R'+R(H'-N')p}{R\sqrt{(H'-N')p(1-p)}})< \beta \leftrightarrow {}$

% $\Phi(1-\alpha)\frac{N'R'+R(H'-N')p}{R\sqrt{(H'-N')p(1-p)}}>\Phi^{-1}$

\noindent $\Phi^{-1}(\beta)=z_{\beta}$, which is the $\beta$-quantile of the normal distribution, which is a constant given $\beta$. Expanding and collecting terms in \(N^{\prime}\) yields the quadratic inequality
\begin{equation}
    a'\,{N^{\prime}}^{2} \;+\; b'\,N^{\prime} \;+\; c' \;>\; 0,
\end{equation}
where
\begin{align*}
    a' \;&=\; \frac{(1-\alpha)^{2}\,(Rp-R^{\prime})^{2}}{R^2},\\
    b' \;&=\; -\frac{p}{R}[2(1-\alpha)^{2}\,H^{\prime}\,(Rp-R^{\prime}) \;+\; z_\beta^2 R^{2}p(1-p)],\\
    c' \;&=\; (1-\alpha)^{2}\,p^{2}{H^{\prime}}^{2} \;-\; z_\beta^2p(1-p)\,H^{\prime},
\end{align*}
% \textcolor{red}{You've used $a,b,c$, so here perhaps $a',b',c'$.}
similar to the previous case. Assume $a'>0$ (i.e., $R'\neq Rp$). We focus on facilities that lack sufficient hash capacity to satisfy Equation (\ref{eq:stochastic_without_pool}), so $H' < \frac{z_\beta^2(1-p)}{(1-\alpha)^2p)}$. This condition is equivalent to $c'<0$, implying that the smaller root is negative. Consequently, the minimal $N'$ that satisfies the constraint is the larger root, clipped at zero:
\begin{equation}
    N^{\prime}_{\min}
    \;=\;
    max(0,\frac{-\,b' \;+\; \sqrt{\,b'^{2}-4a'c'\,}}{2a'}).
\end{equation}

\begin{figure}[t]
    \centering
    \includegraphics[width=1\linewidth]{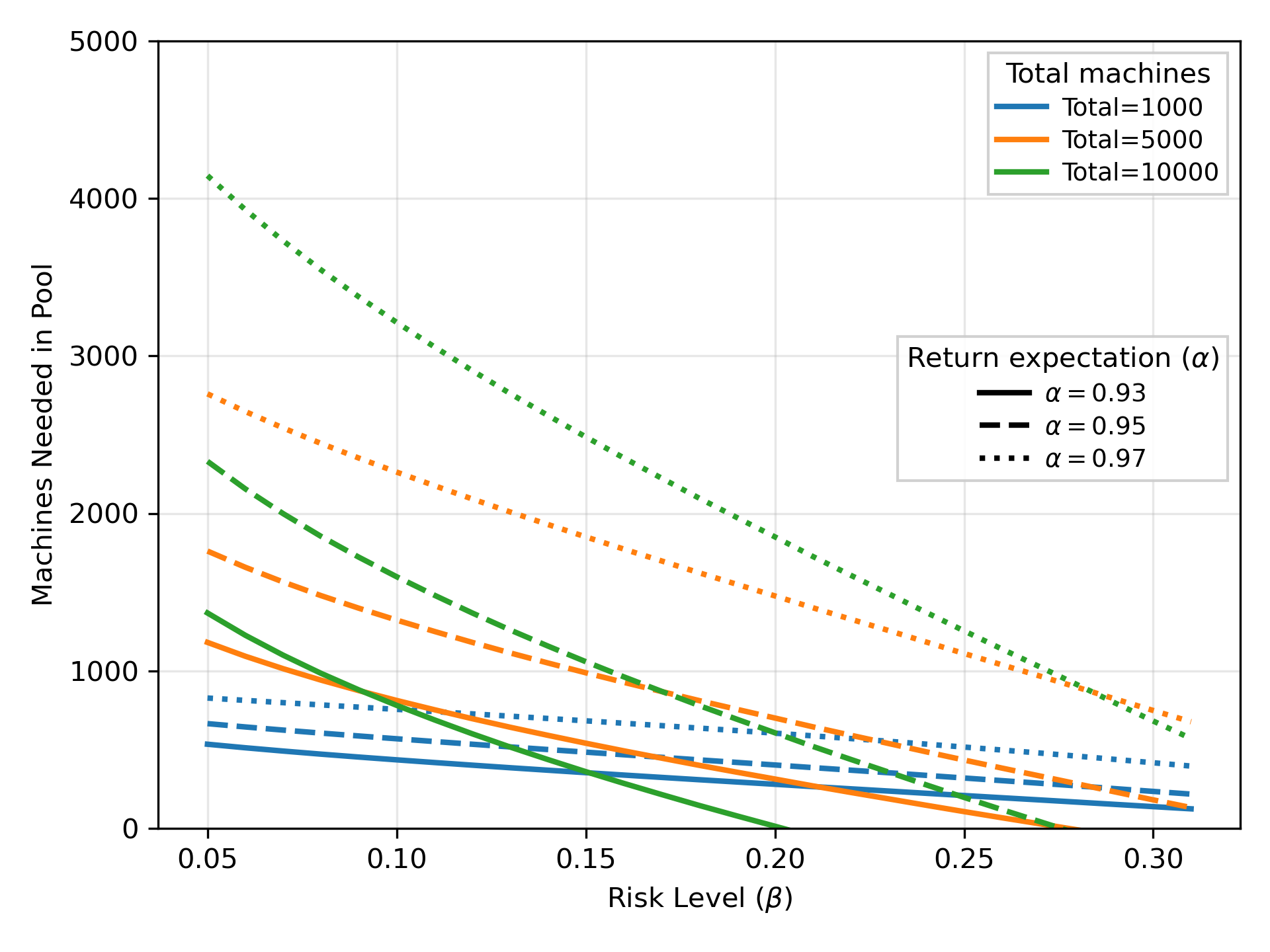}
    \caption{Minimum machines needed to join the mining pool to meet quantile-based risk target, $\beta$, with different return expectation level, $\alpha$, and size of mining facilities.}
    \label{fig:normal_with_pool_alpha}
\end{figure}

\begin{figure}[t]
    \centering
    \includegraphics[width=1\linewidth]{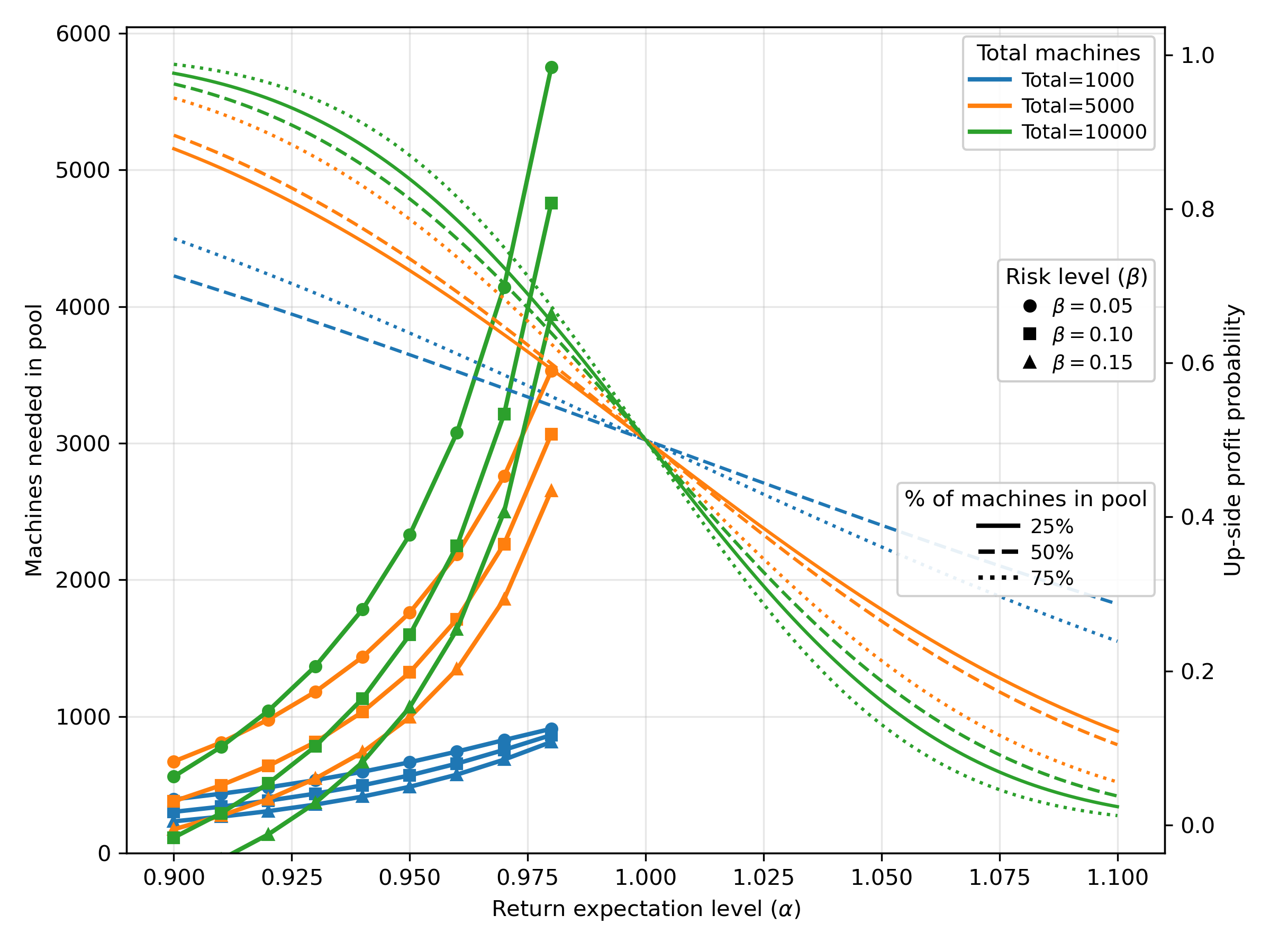}
    \caption{
    Minimum number of machines required to join the mining pool to meet a relative return target $\alpha$ at risk level $\beta$ (solid lines, left axis), and probability that a fleet with total machine $M'$ with fraction of machines allocated to the pool to achieve at least $\alpha$ times its expected profit (dashed lines, right axis).}
    \label{fig:normal_with_pool_beta}
\end{figure}

We thus obtain the approximate minimum number of machines $M'$ from $N'_{min}$, that a facility must allocate to a mining pool to achieve a revenue floor, $\alpha$, at risk tolerance $\beta$ (across facility sizes).
Fig.~\ref{fig:normal_with_pool_alpha} plots $M'$ for a fixed $\alpha$ as $\beta$ varies, while Fig.~\ref{fig:normal_with_pool_beta} plots $M'$ for a fixed $\beta$ as $\alpha$ varies. These results align with the CV case: Pooling tends to benefit smaller facilities (often requiring higher hash-rate penetration), whereas larger facilities typically need pooling only under stringent revenue targets and risk constraints.

% as shown in Fig.~\ref{fig:normal_with_pool_alpha}. The number of machines, $M'$, required to achieve a fixed risk level, $\beta$,
% with different return expectation levels $\alpha$ is shown in Fig.~\ref{fig:normal_with_pool_beta}. These results align with the conclusion from the CV case: Mining pools are more beneficial for smaller facilities but require higher hash-rate penetration, whereas large facilities typically need pools only when revenue targets and risk controls are strict.

Without the normal distribution approximation, finding the minimum hash capability can be described as an optimization problem:
\begin{equation}
    \begin{aligned}
    &min_{N'}N'\\
    &s.t. \sum_{i=1}^{\alpha(H'-N')p+\frac{(\alpha -1)N'R'}{R}} C(H'-N',i)&p^i(1-p)^{H'-N'-i} < \beta.
    \end{aligned}
\end{equation}
The solution gives us the minimum hash rate, $N_{min}^{'}$, required to join the pool to achieve the risk level and profit variance goals, from which we find the number of machines, $M'$, required to reach the goal.

\vspace{0.2cm}
\noindent\textbf{Example 4.} Take $\beta$ = 0.05, $\alpha$ = 0.95, and $H’$ = 10000, the S19 machines’ hash capability. Assume that the per-block reward without pooling is $R=6.25$ BTC, and let $R'$ denote the expected per-hash reward under pool participation. As a rough estimate for $R'$, Riot Platforms reported mining 5,554 BTC in 2022, with 88,556 miners \citep{riot2022sec}. Assuming these are predominantly S19-class machines at 110 TH/s, this implies:

\[R'= \frac{5,554 \ \text{BTC}}{(88556)* (110 \ \text{TH/s})*(365*86400\text{s})} \approx 1.78e^{-22} \ \frac{\text{BTC}}{\text{Hash}}.\]
Under these parameters, the optimal $N_{min}^{'}$ will be around $7.6e^{24}$, which is equivalent to the yearly mining capability of 2191 S19 machines with a mining efficiency of 110 TH/s. More
results are shown in Table \ref{tab:normal_with_pool_opt}. This result agrees with the normal approximation method shown above.

\begin{table}[htp]
    \centering
  \caption{Minimum machines to meet $(\alpha,\beta)$ targets for different size facilities (with mining pool).}
  \label{tab:normal_with_pool_opt}
\begin{tabular}{ |l|l| }

 \hline
  & $M'_{min}$ \\ 
  \hline
 $\# machine = 10000,\alpha = 0.95, \beta = 0.05 $ & 2191 \\ 
  \hline
 $\#  machine = 5000,\alpha = 0.95, \beta = 0.05 $ & 1671 \\ 
  \hline
 $\# machine = 1000,\alpha = 0.95, \beta = 0.05 $ & 619 \\ 
 \hline
\end{tabular}
\end{table}

Similar to the direct mining case, the normal approximation shown in (\ref{eq:normal_pool}) can also be used to determine the probability of achieving different levels of expected net profit under mining-pool participation. Denote the expected total reward when \(N'\) units of hash rate participate in
the pool (with deterministic reward of \(R'\) per unit of hash) and the
remaining \(H' - N'\) units mine directly. For a given relative return factor \(\alpha > 0\), the probability of exceeding \(\alpha\) times this expectation is 
\[
\Pr\bigl(\pi' \ge \alpha \mu_{\pi'}\bigr)
= \Pr\bigl(X' \ge \alpha(H'-N')p + \tfrac{(\alpha-1)N'R'}{R}\bigr),
\]

% we first approximate the underlying binomial distribution by a normal distribution and reuse the normal approximation defined in equation~(\ref{eq:normal_pool}).

% To quantify the expected level of return, we introduce a relative return factor $\alpha \in (0,\infty)$, which scales the expected reward. Let
% \[
% \mu = N'R' + R(H' - N')p
% \]
% denote the expected total reward when $N'$ units of hash rate participate in the pool (with deterministic reward pf $R'$ per hash) and the remaining $H' - N'$ units mine directly with success probability of $p$ and block reward of $R$. Given $N'$, $H'$, $R'$, $R$, and $p$, we are interested in the probability that the realized reward exceeds $\alpha$ times this expectation. This is equivalent to computing
% \[
% \Pr\bigl(X \ge \alpha \mu\bigr)
% = \Pr\bigl(X \ge \alpha(N'R' + R(H' - N')p)\bigr),
% \]
\noindent which is equivalent to
\[
1 - \Pr\bigl(X < \alpha(H'-N')p + \tfrac{(\alpha-1)N'R'}{R}\bigr).
\]
According to the normal approximation in Equation (\ref{eq:normal_prob_pool}), we obtain
% \[
% \begin{aligned}
% \Pr\bigl(X < \alpha(N'R' + R(H' - N')p)\bigr)
% &\approx\\
% \Big(\Phi\!\left((\alpha - 1)\right.\frac{N'R' + R(H' - N')p}{R\sqrt{(H' - N')p(1-p)}}\Big).
% \end{aligned}
% \]
% Therefore,
\begin{equation}
\label{net_profit_pool}
\begin{aligned}
\Pr\bigl(X &\geq \alpha(H'-N')p + \tfrac{(\alpha-1)N'R'}{R}\bigr)
\\ &\approx
1-\Phi\ \Biggl(
  \frac{(\alpha-1)\bigl[(H'-N')p + \tfrac{N'R'}{R}\bigr]}
       {\sqrt{(H'-N')p(1-p)}}
\Biggr).
\end{aligned}
\end{equation}

Once we know the probability of achieving different levels of expected return, the probability distribution of net profit follows directly, because the revenue from the $N'$ hashes in the pool is deterministic, and the electricity cost, $C_e$, is also deterministic for a given fleet size and time horizon.

\vspace{0.2cm}
\noindent\textbf{Example 5.}
Using the calibration results in Section~\ref{sec:calibration}, suppose the facility has a total hash capability, $H'$, corresponding to $1000$ S19 machines mining continuously for one year, and that $250 \ (25\%)$  of these machines join the pool while the remaining $750 \ (75\%)$ machines mine directly. According to Section~\ref{sec:calibration}, the $750$ machines mining without a pool generate expected revenue
\[
750\times \$2{,}962.95=\$2{,}222{,}212.50.
\]
The $250$ machines mining within the pool generate deterministic revenue (based on the bitcoin price in Section~\ref{sec:calibration} and the pool reward estimate in Section~\ref{sec:oppotunity_cost}):
\[
250\times \$42{,}265/\mathrm{BTC}\times0.062~\mathrm{BTC}
=\$655{,}107.50.
\]
Therefore, the total expected revenue under this partial pool strategy is
\[
\$2{,}222{,}212.50+\$655{,}107.50=\$2{,}877{,}320.
\]

Compared to direct mining, this partial pool strategy reduces the expected profit by approximately \$85{,}630. This illustrates the trade-off that motivates our risk-control analysis with mining pools: Although the pool provides an insurance-like effect by smoothing revenue, we still prefer to allocate as little hash rate as necessary to the pool to retain higher upside while keeping risk at an acceptable level.

Setting $\alpha = 1.1$, the probability that the realized revenue exceeds
\[
1.1\times \$2{,}877{,}320=\$3{,}165{,}052
\]
is approximately $35.36\%$, obtained via Equation (\ref{net_profit_pool}). Furthermore, subtracting the deterministic electricity cost, $C_e$, calibrated in Section~\ref{sec:calibration}, where
\[
C_e = 1000\times \$2{,}514.35=\$2{,}514{,}350,
\]
the probability that the net profit exceeds
\[
\$3{,}165{,}052-\$2{,}514{,}350=\$650{,}702
\]
is also $35.36\%$. Additional results for other values of $\alpha$ are summarized in Fig.~\ref{fig:normal_with_pool_beta}. As shown in Fig.~\ref{fig:normal_with_pool_beta}, mining-pool participation reduces the flexibility of returns. For larger facilities and higher pool shares, the probability curve becomes steeper around $\alpha = 1$, indicating a low chance of realizing net profit far above the expectation, but also a low chance of substantial shortfalls. Compared with direct mining, pool participation stabilizes profit at the cost of a lower expected net profit and a reduced probability of achieving a realized profit significantly above the expectation.

Mining pool participation can also be interpreted through the lens of operational hedging. Corporate risk-management theory views hedging as a way to reduce costly cash-flow volatility and coordinate risk management with investment and financing decisions~\citep{smith1985determinants}. Beyond financial contracts, firms can also hedge through operational decisions, including production flexibility, sourcing choices, capacity allocation, and geographic diversification~\citep{boyabatli2004operational}. In our setting, the operational decision is the allocation of physical hash-rate capacity between two revenue-generating modes: direct mining and pool mining. Direct mining preserves the full expected block reward but exposes the facility to block-discovery variance, while pool mining converts part of this stochastic payoff into a lower-variance payout at the cost of fees or reduced expected return. Thus, the fraction of hash capacity allocated to the pool, $\lambda=N'/H'$, can be viewed as an operational hedge ratio. The opportunity cost of hedging is the expected-revenue gap between direct mining and pool participation, while the benefit is the reduction in payoff variance and downside tail risk. Under this interpretation, the risk-control problem in this section characterizes the miner's optimal operational hedge: the minimum pool allocation required to satisfy a given CV or quantile-based risk constraint while preserving as much direct-mining upside as possible.

\section{Grid implications and policy discussion}
\label{sec:policy}
\subsection{Curtailment value and grid-relevant load response}

\begingroup
Curtailment refers to temporarily reducing electricity consumption by shutting down or throttling mining machines, which can resume operation when conditions improve. Under the S19 calibration, one machine draws 3.245 kW and consumes approximately 28.43 MWh annually. Dividing its expected annual revenue of \$2,962.95 by this consumption gives the gross break-even electricity price
\[
p_{e,0}^{*}=\frac{2,962.95}{28.43}\approx \$104.2/\mathrm{MWh}.
\]
A miner exposed to wholesale electricity prices therefore prefers to operate when $p_e<p_{e,0}^{*}$ and to curtail when $p_e\geq p_{e,0}^{*}$. Section~2.2 assumes an electricity price of \$0.0885/kWh, equivalent to \$88.5/MWh. Treating this as an avoidable fixed energy charge gives a net continuation margin of $104.2-88.5=\$15.7$/MWh. This \$15.7/MWh value is relevant only when the miner receives a separate demand-response payment: if the electricity charge is avoided during curtailment, compensation above \$15.7/MWh makes curtailment preferable. The relevant threshold for a miner directly exposed to wholesale prices remains \$104.2/MWh.

The effects of parameter shocks can be calculated from
\begin{equation}
p_e^{*}
=
p_{e,0}^{*}
\left(\frac{P_{\mathrm{BTC}}}{P_{\mathrm{BTC},0}}\right)
\left(\frac{R}{R_0}\right)
\left(\frac{\eta_{e,0}}{\eta_e}\right),
\label{eq:break_even_sensitivity}
\end{equation}

where \(p_{e,0}=\$104.2/\mathrm{MWh}\) is the baseline break-even electricity price. The baseline parameters used in Section~2 are \(P_{\mathrm{BTC},0}=\$42{,}265/\mathrm{BTC}\), \(R_0=6.25\) BTC per block, and \(\eta_{e,0}=29.5\) J/TH. The corresponding variables \(P_{\mathrm{BTC}}\), \(R\), and \(\eta_e\) denote the Bitcoin price, block reward, and hardware energy consumption under the alternative scenario. Holding other parameters constant, a 20\% decrease in the Bitcoin price lowers the threshold to $104.2\times0.8=\$83.4$/MWh, while a 20\% increase raises it to \$125.1/MWh. Halving the block reward lowers it to \$52.1/MWh; and improving hardware efficiency from 29.5 to 21.1 J/TH raises it to $104.2\times29.5/21.1=\$145.7$/MWh.

Mining-pool participation also affects the economics of curtailment. In Example~5, allocating 25\% of the 1000-machine fleet to the mining pool reduces expected annual revenue from \$2,962,950 under direct mining to \$2,877,320, while annual electricity consumption remains approximately 28,426.2 MWh. Consequently, the gross break-even electricity price decreases from \$104.2/MWh to approximately \$101.2/MWh, a reduction of about 2.9\%. Under the baseline electricity rate of \$88.5/MWh, and assuming that this electricity charge is avoided during curtailment, the net mining margin—and hence the minimum separate curtailment payment—decreases from \$15.7/MWh under direct mining to \$12.7/MWh under the partial-pool strategy. This represents a reduction of approximately \((15.7-12.7)/15.7=19.1\%\). Thus, the 19.1\% reduction specifically measures the effect of the 25\% pool-allocation strategy on the expected net income forgone when mining load is curtailed.

In ERCOT, qualified Controllable Load Resources can submit price-responsive bids and be dispatched downward when wholesale prices reach their bid levels~\citep{ercot_load_resources}. The model therefore provides an economic basis for constructing these bids and demonstrates why mining flexibility should be modeled as state-dependent rather than as a fixed quantity of curtailable load.

\subsection{Implications for grid operators and regulators}
Mining load should be modeled as state-dependent demand rather than as fixed consumption or a fixed quantity of flexibility. ERCOT already incorporates Bitcoin-market conditions into resource-adequacy forecasting. Its March 2025 assessment included a 2,934-MW adjustment for large flexible loads using an estimated mining break-even price of \$80.26/MWh and assumed that facilities would reduce consumption to approximately 5\% of maximum load above this threshold~\citep{ercot2025mora}. The proposed framework generalizes this approach by allowing the break-even price to vary with Bitcoin price, block reward, difficulty, hardware efficiency, and pool participation.

Grid operators should forecast expected mining consumption separately from additional curtailable load. As shown in Section~5.1, a facility may operate under the baseline state but curtail following a 20\% decline in the Bitcoin price. Moreover, mining load that is already offline cannot provide further demand response. Updating load forecasts, bids, and curtailment procurement using contemporaneous mining conditions can therefore reduce both forecasting errors and overestimation of available flexibility.

Regulators should similarly distinguish nameplate demand, expected coincident consumption, and verified price-responsive load reduction. Texas already requires certain mining facilities with more than 75 MW of load and at least 10\% interruptible load to report their anticipated peak demand and interruptible share~\citep{puct2024vcm}. Because an interruptible percentage does not identify the price at which curtailment occurs, these disclosures could be supplemented with price-responsive demand schedules. Consistent with NERC recommendations~\citep{nerc2026largeloads}, flexibility should receive planning or interconnection credit only when it is contractually enforceable and operationally verified.
\endgroup

\section{Conclusion and future work}
\label{sec:conclusion}

In this paper, we developed a unified, ex ante statistical framework for bitcoin mining that derives economic and energy quantities directly from the underlying proof-of-work process. By modeling each hash as a Bernoulli trial with success probability determined by network difficulty, we obtained closed-form expressions for expected revenue, costs, and net profit per unit of hash rate. Building on this probabilistic core, we analyzed methods for controlling downside risk and characterizing upside profit potential for both direct and pool mining, and calibrated the framework using recent market data.

% The framework makes the trade-offs between profitability and risk explicit. For direct mining, we showed how coefficient-of-variation and $\beta$-quantile constraints can be used to bound downside risk or enforce a desired confidence level on net profit. For pool mining, we quantified the opportunity cost of participation and demonstrated how increasing pool share reduces variance but also dampens upside potential. Our net-profit analysis, including the probability of exceeding an $\alpha$-multiple of expected profit for different fleet sizes and pool-participation strategies, illustrates how miners can choose operating points that balance higher expected returns against tighter risk control.

From a practical perspective, the results provide miners with transparent formulas and graphical tools for fleet sizing, pool-participation decisions, and planning under uncertainty. For system operators and planners, the same probabilistic framework offers a miner's decision-aware way to translate prices and incentives into expected mining load and flexibility, supporting the design of demand-response and ancillary-service programs that treat mining as a controllable load.

Several limitations point to future work. Our current revenue model simplifies fee dynamics and pool payout rules and abstracts from location-specific constraints such as curtailment, network congestion, and carbon intensity. Future extensions include incorporating richer fee and transaction-revenue processes, explicitly modeling endogenous participation under real-time prices, and embedding the mining model in a joint equilibrium with grid conditions and locational signals. 

Overall, by moving from deterministic snapshot proxies to a coherent probabilistic framework, this paper provides a durable foundation for analyzing mining profitability and risk, and for designing policies that responsibly harness the flexibility of large-scale mining loads.

\bibliographystyle{elsarticle-harv} 
\bibliography{references}

@article{krause2018quantification,
  title   = {Quantification of energy and carbon costs for mining cryptocurrencies},
  author  = {Krause, Max J. and Tolaymat, Thabet},
  journal = {Nature Sustainability},
  volume  = {1},
  number  = {11},
  pages   = {711--718},
  year    = {2018},
  publisher = {Nature Publishing Group UK London}
}

@article{menati2023high,
  title   = {High-resolution modeling and analysis of cryptocurrency mining's impact on power grids: Carbon footprint, reliability, and electricity price},
  author  = {Menati, Ali and Zheng, Xiangtian and Lee, Kiyeob and Shi, Ranyu and Du, Pengwei and Singh, Chanan and Xie, Le},
  journal = {Advances in Applied Energy},
  volume  = {10},
  pages   = {100136},
  year    = {2023},
  publisher = {Elsevier}
}

@techreport{wade2025electricity,
  title        = {Electricity Grid Impacts of Rising Demand from Data Centers and Cryptocurrency Mining Operations},
  author       = {Wade, Cameron and Blackhurst, Mike and DeCarolis, Joe and de Queiroz, Anderson and Johnson, Jeremiah and Jaramillo, Paulina},
  institution  = {{Open Energy Outlook Initiative, Scott Institute for Energy Innovation, Carnegie Mellon University}},
  month        = {June},
  year         = {2025},
  url          = {https://energy.cmu.edu/_files/documents/electricity-grid-impacts-of-rising-demand-from-data-centers-and-cryptocurrency-mining-operations.pdf},
  note         = {White paper}
}

@article{menati2023modeling,
  title   = {Modeling and analysis of utilizing cryptocurrency mining for demand flexibility in electric energy systems: A synthetic {Texas} grid case study},
  author  = {Menati, Ali and Lee, Kiyeob and Xie, Le},
  journal = {IEEE Transactions on Energy Markets, Policy and Regulation},
  volume  = {1},
  number  = {1},
  pages   = {1--10},
  year    = {2023},
  publisher = {IEEE}
}

@article{majumder2024econometric,
  title   = {An Econometric Analysis of Large Flexible Cryptocurrency-Mining Consumers in Electricity Markets},
  author  = {Majumder, Subir and Aravena, Ignacio and Xie, Le},
  journal = {arXiv preprint arXiv:2408.12014},
  year    = {2024}
}

@unpublished{menati2024optimization,
  title   = {Optimization of Cryptocurrency Mining Demand for Ancillary Services in Electricity Markets},
  author  = {Menati, Ali and Cai, Yuting and El Helou, Rayan and Tian, Chao and Xie, Le},
  year    = {2024},
  note    = {Manuscript}
}

@techreport{campecino2021portfolio,
  title       = {Portfolio Theory and Risk Analysis Using Coefficient of Variation: An Alternative to the Modern Portfolio Theory},
  author      = {Campeci{\~n}o, J. O.},
  institution = {Michigan State University},
  number      = {2109},
  year        = {2021}
}

@article{rogers2023bitcoin,
  title   = {{Bitcoin} equilibrium dynamics: a long-term approach},
  author  = {Rogers, Jack R.},
  journal = {Frontiers in Blockchain},
  volume  = {6},
  year    = {2023},
  url     = {https://www.frontiersin.org/articles/10.3389/fbloc.2023.1226892},
  doi     = {10.3389/fbloc.2023.1226892},
  issn    = {2624-7852}
}

@misc{riot2022sec,
  author = {Riot Platforms, Inc.},
  title  = {Annual Report Pursuant to Section 13 or 15(d) of the Securities Exchange Act of 1934},
  year   = {2023},
  url    = {https://www.sec.gov/ix?doc=/Archives/edgar/data/0001167419/000155837023002704/riot-20221231x10k.htm},
  note   = {Form 10-K}
}

@misc{mara2022sec,
  author = {{Marathon Digital Holdings, Inc.}},
  title  = {Annual Report Pursuant to Section 13 or 15(d) of the Securities Exchange Act of 1934},
  year   = {2023},
  url    = {https://www.sec.gov/ix?doc=/Archives/edgar/data/0001507605/000149315223007879/form10-k.htm},
  note   = {Form 10-K}
}

@article{neumueller2025cambridge,
  title={Cambridge Digital Mining Industry Report: Global Operations, Sentiment, and Energy Use},
  author={Neumueller, Alexander and Pieters, Gina C and Mohaddes, Kamiar and Rousseau, Valentin and Zhang, Bryan Zheng},
  journal={Sentiment, and Energy Use (April 28, 2025)},
  year={2025}
}

@online{coinwarz_difficulty_2025,
  title   = {Bitcoin Difficulty Chart -- BTC Difficulty},
  author  = {{CoinWarz}},
  url     = {https://www.coinwarz.com/mining/bitcoin/difficulty-chart},
  urldate = {2025-11-06}
}

@online{bitmain_s19pro_specs,
  author = {{BITMAIN}},
  title  = {S19 Pro Specifications},
  year   = {2021},
  note   = {Updated May 27, 2021},
  url    = {https://support.bitmain.com/hc/en-us/articles/900000261726-S19-Pro-Specifications},
  urldate= {2025-11-06}
}

@misc{statmuse_bitcoin_2023,
  author       = {StatMuse},
  title        = {Bitcoin price history 2023},
  howpublished = {\url{https://www.statmuse.com/money/ask/bitcoin-price-history-2023?utm_source=chatgpt.com}},
  note         = {Accessed: November 17, 2025},
  year         = {2023}
}

@misc{riot2025jan,
  author       = {{Riot Platforms}},
  title        = {{Riot Announces January 2025 Production and Operations Updates}},
  year         = {2025},
  month        = feb,
  day          = {4},
  howpublished = {\url{https://www.riotplatforms.com/riot-announces-january-2025-production-and-operations-updates/}},
  note         = {Accessed: 2026-05-31}
}

@article{smith1985determinants,
  title={The determinants of firms' hedging policies},
  author={Smith, Clifford W and Stulz, Rene M},
  journal={Journal of financial and quantitative analysis},
  volume={20},
  number={4},
  pages={391--405},
  year={1985},
  publisher={Cambridge University Press}
}

@article{boyabatli2004operational,
  title={Operational hedging: A review with discussion},
  author={Boyabatli, Onur and Toktay, L Beril},
  year={2004}
}

@misc{harper2022hashrate,
author = {Harper, Colin},
title = {{Hashrate Index Roundup 3/13/22}},
year = {2022},
month = mar,
day = {13},
howpublished = {\url{https://hashrateindex.com/blog/hashrate-index-roundup-3-13-22/}},
note = {Hashrate Index}
}

@misc{cleanspark2024dec,
  author       = {{CleanSpark, Inc.}},
  title        = {{CleanSpark Exceeds 2024 Year-End Guidance of 37 EH/s and Accelerates 2025 Guidance}},
  year         = {2024},
  month        = {December},
  day          = {23},
  howpublished = {\url{https://investors.cleanspark.com/news/news-details/2024/CleanSpark-Exceeds-2024-Year-End-Guidance-of-37-EHs-and-Accelerates-2025-Guidance/default.aspx}},
  note         = {Accessed: August 8, 2026}
}

@misc{mara2025dec,
  author       = {{MARA Holdings, Inc.}},
  title        = {{MARA Announces Bitcoin Production and Mining Operation Updates for December 2024}},
  year         = {2025},
  month        = {January},
  day          = {3},
  howpublished = {\url{https://ir.mara.com/news-events/press-releases/detail/1385/mara-announces-bitcoin-production-and-mining-operation-updates-fordecember-2024}},
  note         = {Accessed: August 8, 2026}
}

@misc{eia2025residential,
  author       = {{U.S. Energy Information Administration}},
  title        = {{Residential Electric Bills in Hawaii and Connecticut Are Twice Those in New Mexico, Utah}},
  year         = {2025},
  month        = {May},
  day          = {12},
  howpublished = {\url{https://www.eia.gov/todayinenergy/detail.php?id=65244}},
  note         = {Accessed: August 8, 2026}
}

@techreport{ercot2025mora,
  author      = {{Electric Reliability Council of Texas}},
  title       = {Monthly Outlook for Resource Adequacy: March 2025},
  institution = {ERCOT},
  year        = {2025},
  month       = jan,
  url         = {https://www.ercot.com/files/docs/2025/01/03/MORA_March2025.pdf},
  note        = {Accessed: August 8, 2026}
}

@misc{puct2024vcm,
  author = {{Public Utility Commission of Texas}},
  title  = {Registration of Virtual Currency Mining Facilities,
            16 Texas Administrative Code Section 25.114},
  year   = {2024},
  url    = {https://ftp.puc.texas.gov/public/puct-info/agency/rulesnlaws/subrules/electric/25.114/25.114.pdf},
  note   = {Effective December 11, 2024; accessed August 8, 2026}
}

@techreport{nerc2026largeloads,
  author      = {{North American Electric Reliability Corporation}},
  title       = {Assessment of Gaps in Existing Practices, Requirements,
                 and Reliability Standards for Emerging Large Loads},
  institution = {NERC},
  year        = {2026},
  month       = mar,
  url         = {https://www.nerc.com/globalassets/our-work/guidelines/reliability/white-paper---assessment-of-gaps.pdf},
  note        = {Accessed: August 8, 2026}
}

@misc{ercot_load_resources,
  author = {{Electric Reliability Council of Texas}},
  title  = {Load Resource Participation in the ERCOT Markets},
  url    = {https://www.ercot.com/services/programs/load/laar},
  note   = {Accessed: August 8, 2026}
}

\appendix
% \section{Appendix: The difficulty Adjustment}
% \label{sec11}
% Bitcoin targets adjust every two weeks according to the difficulty adjustment. This ensures that block times remain on average 10 minutes apart. There are 26 difficulty adjustment periods per year. The current target can be calculated from within each block header (embedded inside the bits field). Therefore, we can extract the target from the Bitcoin blockchain. 
% We can, therefore, calculate the target for each of the 26 difficulty adjustment periods per year. With this data we can then calculate the expected value per tera hash per second, using EV* in each difficulty adjustment period during the year. Let
% \begin{align*}
% H_t=\frac{2^{256}}{\tau_tB_t}
% \end{align*}
% be the expected number of hashes needed to mine 1 BTC, given target t and block subsidy Bt in difficulty adjustment period t, ranging from t=1 to t=26. 
% For convenience, an “\pi-miner” is a bitcoin miner that hashes \pi terahash per second. For example, the ASIC-S19 miner by Bitmain is a miner of hash rate \pi=110. Therefore, an \pi-miner will generate an expected value of 
% \begin{align*}
% \begin{split}
% EV(\pi-miner) = \\\pi(EV_t)(60 \times 60 \times 24 \times 14)
% \end{split}
% \end{align*}
% over each two-week period.
% \end{multicols}

\section{Machine efficiency and network dominance}

\begin{table}[h]
  \centering
  \caption{Bitcoin mining machine efficiency and network dominance.}
  \begin{tabular}{lcc}
    \hline
    Model & Network dominance (\%) & Efficiency (J/GH) \\
    \hline
    S19j Pro & 34.31 & 0.031 \\
    S19      & 28.10 & 0.034 \\
    S19 XP   & 11.33 & 0.022 \\
    M20S     & 7.00  & 0.049 \\
    M32      & 5.18  & 0.054 \\
    1246     & 4.74  & 0.036 \\
    M50      & 2.59  & 0.029 \\
    1066     & 2.54  & 0.065 \\
    S9       & 2.40  & 0.093 \\
    S17      & 1.28  & 0.045 \\
    E12+     & 0.53  & 0.050 \\
    \hline
  \end{tabular}
\label{tab:mining-machines}

\end{table}
\end{document}